\documentclass{article}

\usepackage{amsfonts,amsmath,amsthm,color,times,fullpage} 
\usepackage{graphicx}

\newtheorem{theorem}{Theorem}
 
\newtheorem{claim}[theorem]{Claim} 
\newtheorem{corollary}[theorem]{Corollary}

\newtheorem{definition}[theorem]{Definition}
 
\newtheorem{lemma}[theorem]{Lemma} 
\newtheorem{remark}[theorem]{Remark}  

\newcommand{\whp} {w.h.p.}
\newcommand{\whitening}{coarsening}
\newcommand{\Whitening}{Coarsening}
\newcommand{\ie}{i.e.,\ }
\newcommand{\eg}{e.g.,\ }

\newcommand{\ex}{\mathbb{E}} 
\newcommand{\parta}{\alpha}
\newcommand{\Makis}{\Delta}

\renewcommand{\a}{\alpha}
\newcommand{\g}{\gamma}

\begin{document}

\title{On the Solution-Space Geometry of\\
Random Constraint Satisfaction Problems}

\author{Dimitris Achlioptas \\
  Department of Computer Science,
  University of California Santa Cruz \\
  {\tt optas@cs.ucsc.edu}
  \and
  Federico Ricci-Tersenghi \\
  Physics Department, University of Rome ``La Sapienza''\\
  {\tt federico.ricci@roma1.infn.it}
}

\date{\empty}

\maketitle

\begin{abstract}
For a large number of random constraint satisfaction problems, such as
random k-SAT and random graph and hypergraph coloring, there are very
good estimates of the largest constraint density for which solutions
exist. Yet, all known polynomial-time algorithms for these problems
fail to find solutions even at much lower densities. To understand the
origin of this gap we study how the structure of the space of
solutions evolves in such problems as constraints are added. In
particular, we prove that much before solutions disappear, they
organize into an exponential number of clusters, each of which is
relatively small and far apart from all other clusters. Moreover,
inside each cluster most variables are frozen, i.e., take only one
value. The existence of such frozen variables gives a satisfying
intuitive explanation for the failure of the polynomial-time
algorithms analyzed so far. At the same time, our results establish
rigorously one of the two main hypotheses underlying Survey
Propagation, a heuristic introduced by physicists in recent years that
appears to perform extraordinarily well on random constraint
satisfaction problems.
\end{abstract}
 
\section{Introduction}

For a number of random Constraint Satisfaction Problems (CSP), by now
very good estimates are available for the largest constraint density
(ratio of constraints to variables) for which typical problems have
solutions. For example~\cite{AN}, a random graph of average degree $d$
is with high probability~\footnote{We will say that a sequence of
events ${\mathcal E}_n$ occurs with high probability (\whp) if
$\lim_{n \to \infty} \Pr[{\mathcal E}_n] =1$.} $k$-colorable if $d <
(2k-2) \ln (k-1)$, but \whp\ non-$k$-colorable if $d > (2k-1) \ln
k$. This implies that for every $d>0$, \whp\ the chromatic number of a
random graph with average degree $d$ is either $k_d$ or $k_d+1$, where
$k_d$ is the smallest integer $k$ such that $d < 2k \ln k$.

Algorithmically, it is very easy to get a factor-2 approximation for
the graph coloring problem on random graphs. The algorithm
```repeatedly pick a random vertex and assign it a random available
color" will \whp\ succeed in coloring a random graph of average degree
$d$ if originally each vertex has $2 k_d$ available
colors. Alternatively, $k$ colors suffice when $d < k \ln k$. In spite
of significant efforts over the last 30 years, no improvement has been
made over this trivial algorithm. Specifically, no polynomial-time
algorithm is known that $k$-colors random graphs of average degree $d
= (1+\epsilon) k \ln k$, for some fixed $\epsilon >0$ and arbitrarily
large $k$.

In the random $k$-SAT problem one asks if a random $k$-CNF formula,
$F_k(n,m)$, with $n$ variables and $m$ clauses is satisfiable. It is
widely believed that the probability that such a formula is
satisfiable exhibits a sharp threshold. Specifically, the {\em
Satisfiability Threshold Conjecture\/} asserts that $r_k = r_k^*$ for
all $k \geq 3$, where
\begin{eqnarray*}
  r_k   & \equiv & \sup\{r : F_k(n,rn)\mbox{ is satisfiable    \whp}\}
  \enspace ,\\
  r_k^* & \equiv & \inf\{r : F_k(n,rn)\mbox{ is unsatisfiable  \whp}\}
  \enspace .
\end{eqnarray*}

It is easy to see that $r_k^* \leq 2^k \ln 2$, since the probability
that at least one assignment satisfies $F_k(n,rn)$ is bounded by $2^n
(1-2^{-k})^{rn}$, a quantity that tends to 0 for $r \geq 2^k \ln
2$. Recently, it was shown that random $k$-CNF formulas have
satisfying assignments for densities very close to this upper
bound~\cite{AP}. Specifically, it was proven that for all $k \geq 3$,
\begin{equation}\label{acp}
  r_k > 2^k \ln 2 - \frac{(k+1)\ln 2 + 3}{2} \enspace .
\end{equation}
As for the $k$-coloring problem, the lower bound on the largest
density for which solutions exist \whp\ is non-constructive, based on
the second moment method. Here, the gap relative to algorithms is ever
greater: no polynomial algorithm is known that finds satisfying
assignments in a random $k$-CNF formula when $r = \omega(k)\,2^k/ k$,
for any function $\omega(k) \to \infty$ (arbitrarily slowly). In
Table~\ref{tab:val}, we illustrate this gap for some small values of
$k$.  For $k=3$, the upper bound on $r_k^*$ comes from \cite{DuBoMa},
while for $k>3$ from~\cite{DuBo,KKKS}.  The best algorithmic lower
bound for $k=3$ is from~\cite{352}, while for $k>3$ it is from
\cite{FrSu}.

\begin{table*}\label{tab:val}
\centering
\[
\begin{array}{c|cccccc}
 k                                        & 3     & 4     & 7     &
10     & 20      & 21 \\ \hline
\mbox{Best known upper bound for $r_k^*$} & 4.508 & 10.23 & 87.88 &
708.94 & 726,817 & 1,453,635 \\
\mbox{Best known lower bound for $r_k$}   & 3.52  & 7.91  & 84.82 &
704.94 & 726,809 & 1,453,626 \\
\mbox{Best known algorithmic lower bound} & 3.52  & 5.54  & 33.23 &
172.65 &  95,263 & 181,453
\end{array}
\]
\end{table*}

Similar results (and gaps) exist for a number of other constraint
satisfaction problems, such as random \mbox{NAE $k$-SAT} and
hypergraph 2-coloring, regular random graph coloring, random Max
$k$-SAT, and others (for example, see~\cite{ANP}). Indeed, this
phenomenon seems to occur in nearly all random CSP in which the
underlying constraint graph is sparse and random, making it natural to
ask if there is a common underlying cause. (The bipartite graphs where
constraints are adjacent to the variables they bind are also known as
factor graphs.)

As it turns out, sparse random CSP have been systematically studied by
physicists in the past few decades under the name ``mean-field diluted
spin-glasses". Spins here are the variables (reflecting the notion
that variables have small, discrete domains), the term glass refers to
the fact that the system has not been allowed to relax to a
configuration in which spins interact in a mutually agreeable way
(reflecting that different constraints prefer different values for the
variables), diluted refers to the fact that the factor graph is sparse
(reflecting that each spin interacts with only a few other spins),
while ``mean field" refers to the fact that the factor graph is
random, \ie there is no underlying geometry mandating the
interactions. The interest in such ``unphysical'' systems is partly
motivated by the fact that in many statistical mechanics problems
where the variables do lie on a lattice such as ${\mathbb{Z}}^d$, the
effect of the underlying geometry vanishes for $d$ sufficiently large
(but finite).

Perhaps more surprising is the fact that in the last few years,
motivated by ideas developed for the study of materials, physicists
have put forward a hypothesis for the origin of the aforementioned
algorithmic gap in random CSP and, most remarkably, a method for
overcoming it. Specifically, M\'{e}zard, Parisi, and
Zecchina~\cite{sp} developed an extremely efficient algorithm, called
Survey Propagation (SP), for finding satisfying assignments of random
formulas in the satisfiable regime. For example, their algorithm
typically finds a satisfying truth assignment of a random 3-CNF
formula with $n=10^6$ variables and $4.25n$ clauses in minutes (and
appears to scale as $O(n \log n)$). No other algorithm practically
solves formulas of such density with $n= 10^4$.

Our original motivation for this work was to see if some of the
physically-motivated ideas underlying SP can be proven mathematically.
More generally, we believe that understanding the geometry of the
space of satisfying truth assignments in random formulas is essential
for understanding the behavior of algorithms on them. This is
particularly true for the case of random-walk type algorithms, which
we view as the first natural class to target armed with such an
understanding and for which very little is known rigorously, with the
notable exception of~\cite{ab}.

We make significant progress towards this goal by proving that already
much below the satisfiability threshold, the set of satisfying
assignments fragments into exponentially many connected components.
These components are relatively small in size, far apart from one
another, and inside each one the majority of variables are ``frozen",
\ie take only one value. As the formula density is increased towards
the threshold, the fraction of frozen variables in each component
increases, causing the connected components to decrease in volume and
grow further apart from one another.

Our results are in perfect agreement with the picture put forward by
the physicists. Moreover, as we discuss below, the existence of frozen
variables provides a good explanation for the origin of the barrier
faced by all analyzed algorithms on random CSP, \ie ``local",
DPLL-type algorithms. Finally, we show that one of the two main
assumptions underlying SP regarding the structure of the set of
solutions is essentially correct. This brings us closer to a rigorous
analysis of SP and answers affirmatively the main open question raised
by Maneva, Mossel and Wainwright in~\cite{elitza}.

Specifically, we prove that for all $k \geq 9$, the connected
components of the set of satisfying assignments of random formulas
have non-trivial cores, as assumed by SP (see
Definition~\ref{core_def}). We point out that it is not clear whether
this is true for small $k$. Indeed, \cite{elitza} gave experimental
evidence that for $k=3$, random formulas do {\em not\/} have
non-trivial cores. As we will see, our methods also give some evidence
in that direction. This gives additional motivation for the
``core-like" objects introduced in~\cite{elitza} whose existence would
relate to the success of SP for small $k$ (we discuss this point
further in Section~\ref{sec:sp_related}).

In the next section we give an informal discussion relating the
performance of DPLL-type algorithms on random formulas to notions such
as Gibbs sampling and long-range correlations. This is meant to
provide intuition for the empirical success of SP and motivate our
results. We emphasize that while both the discussion and the results
are about random \mbox{$k$-SAT}, this is not strictly necessary: our
ideas and proofs are quite generic, and should generalize readily to
many other random CSP, \eg graph coloring.

\subsection{DPLL algorithms, Belief Propagation, and Frozen
Variables}

Given a satisfiable formula $F$ on variables $v_1,v_2,\ldots,v_n$ it
is easy to see that the following simple procedure samples uniformly
from the set of all satisfying assignments of $F$:
\begin{itemize}
\item[] \hspace*{-0.4cm} Start with the input formula $F$
\item[] \hspace*{-0.4cm} 
  For $i=1$ to $n$ do: 
  \begin{enumerate}
  \item
    Compute the fraction, $p_i$, of satisfying assignments of the current
    formula in which $v_i$ takes the value 1.
  \item
    Set $v_i$ to 1 with probability $p_i$ and to 0 otherwise.
  \item
    Simplify the formula.
  \end{enumerate}
\end{itemize}

Clearly, the first step in the loop above is meant only as a thought
experiment. Nevertheless, it is worth making the following two
observations. The first is that if we are only interested in finding
{\em some\/} satisfying assignment, as opposed to sampling a uniformly
random one, then we do not need to compute exact marginals. For
example, if we use the rule of always setting $v_i$ to 1 iff $p_i \geq
1/2$, then it is enough to ensure that if a variable takes the same
value $x$ in {\em all} satisfying assignments, $x$ is the majority
value in its computed marginal. The second observation is that the
order in which we set the variables does not need to be determined a
priori. That is, we can imagine that in each step we compute marginals
for all remaining variables and that for each marginal we have an
associated confidence. To improve our chances of avoiding a fatal
error, we can then set only the variable for which we have highest
confidence.

The above two elementary observations in fact capture all algorithms
that have been analyzed so far on random formulas (and, in fact, most
DPLL-type algorithms used in practice). Observe, for example, that
both the {\em unit-clause\/} and the {\em pure literal\/} heuristics
follow immediately from the above considerations. In the case of
unit-clause, the participation of a variable $v$ in a unit clause $c$
allows us to infer its marginal with perfect confidence and thus
setting $v$ is an ``obvious" choice. In the case of a pure literal
$\ell$, again we can infer with certainty the majority marginal of the
underlying variable $v$ (it is the value that satisfies $\ell$). In
the absence of such obvious choices, all DPLL-type algorithms attempt
to identify a variable whose marginal can be determined with some
confidence. For example, below are the choices made in the absence of
unit clauses and pure literals by some of the algorithms that have
been analyzed on random 3-CNF formulas. In order of increasing
performance:
\begin{itemize}
\item[]{\sc unit-clause}~\cite{mick}: select a random variable and
  assign it a random value.
\item[]{\sc 3-clause majority}~\cite{ChFrGUC}: select a random
  variable and assign it its majority value among the 3-clauses.
\item[]{\sc short-clause}\cite{FrSu}: select a random shortest clause
  $c$, a random variable $v$ in $c$, and set $v$ so as to satisfy $c$.
\item[]{\sc happiest literal~\cite{342}:} satisfy a literal that
  appears in most clauses.
\end{itemize}

Each of the above heuristics attempts to compute marginals based on a
different set of evidence, the content of which ranges from completely
empty~\cite{mick}, to considering all the clauses containing each
variable~\cite{342}. Correspondingly, the largest density for which
these algorithms succeed on random 3-CNF formulas ranges from $8/3$
for~\cite{mick} to 3.42 for~\cite{342}. {\sc Unit-clause}, in fact,
succeeds for every $k$ as long as $r < 2^k/k$ and, as we mentioned
earlier, no algorithm is known to beat this bound
asymptotically. Given that improving upon the empty set of evidence is
rather easy, it is tempting to think that by considering a larger set
of evidence for each variable one can do significantly better. For
example, consider an algorithm ${\cal A}_d$ which computes a marginal
for each variable $v$ based on the clauses that appear in the
depth-$d$ neighborhood of $v$ in the factor graph. One could hope that
as $d$ grows, such an algorithm would do well, perhaps even reach the
satisfiability threshold.

Physicists say it is not so. The hope that local algorithms could do
well on random formulas rests on the presumption that the influence
exerted on a variable $v$ by other variables, diminishes rapidly with
their distance from $v$ in the factor graph. That is, that there are
no ``long range correlations" in random formulas, so that the joint
probability distribution of a random finite subset of the variables
should be, essentially, the product of their marginals.

Unfortunately, the existence of connected components of satisfying
assignments (clusters) with numerous frozen variables can induce
long-range correlations among the variables, eliminating such
hopes. For example, if one considers any fixed set of variables, at
sufficiently large densities their joint behavior over the set of
satisfying assignments is dominated by a small number of connected
components: those in which most of them are frozen, freeing up other
variables to take multiple values and amplify the contribution of that
particular joint collection of values to the variable marginals. In
other words, assuming that the variables in the boundary of a
variable's neighborhood behave independently with respect to the rest
of the formula, can be very far off from the truth.

To overcome the above issue, physicists hypothesized that the above
clustering is the only significant source of long-range
correlations. (Very) roughly speaking, this amounts to modeling each
connected component of satisfying assignments as a subcube that
results by selecting a large fraction of the variables and freezing
them independently at random, while leaving the rest (largely)
free. Our results imply that this simplified view of clusters is not
very far off the truth.

\subsection{Organization}

In the next section we give mathematical statements of our main
results, regarding the existence of exponentially many well-separated
clusters and the existence of frozen variables in every cluster. In
Section~\ref{sec:clustering} we outline the proof of the results on
the existence of clusters and explain their relationship to the work
of Mora, M\'ezard, and Zecchina~\cite{phys_clus_j,phys_clus_l}. In
Section~\ref{sec:frozeness} we provide some background on Survey
Propagation and explain how our results on frozen variables relate to
the implicit hypothesis made by M\'ezard, Parisi, and Zecchina in
their derivation of Survey Propagation~\cite{sp}, and how our results
answer the main open question of Maneva, Mossel and
Wainwright~\cite{elitza}. In Section~\ref{sec:setup} we introduce the
probabilistic setup for our analysis and in
Section~\ref{sec:plantedModel} we discuss how it relates to the case
$k < 8$ and to the ``planted assignment" model. Proving our main
result on the existence of frozen variables boils down to a question
in large deviations developed in Section~\ref{sec:ana} and an
associated multi-dimensional optimization problem, resolved in
Section~\ref{sec:opt}.

\section{Statement of Results}

We first need to introduce some definitions. Throughout, we assume
that we are dealing with a CNF formula $F$, defined over variables
$X=x_1,\ldots,x_n$, and we let $\mathcal{S}(F) \subseteq \{0,1\}^n$
denote the satisfying assignments of $F$.

\begin{definition}\label{def:clus}
The \textbf{diameter} of an arbitrary set $X \subseteq \{0,1\}^{n}$ is
the largest Hamming distance between any two elements of $X$. The
\textbf{distance} between two arbitrary sets $X,Y\subseteq
\{0,1\}^{n}$, is the minimum Hamming distance between any $x\in X$ and
any $y\in Y$. The \textbf{clusters} of a formula $F$ are the connected
components of $\mathcal{S}(F)$ when $x,y \in \{0,1\}^n$ are considered
adjacent if they have Hamming distance 1. A \textbf{cluster-region} is
a non-empty set of clusters.
\end{definition}

Our first set of results is captured by Theorems~\ref{basic} and
\ref{sharp_basic} below, which build upon the work
in~\cite{AP,phys_clus_j,phys_clus_l}. We discuss the
relationship between our results and those
in~\cite{phys_clus_j,phys_clus_l} in Section~\ref{sec:related}.
\begin{theorem}\label{basic}
For every $k \geq 8$, there exists a value of $r<r_{k}$ and constants
$\alpha_{k} < \beta_{k}<1/2$ and $\epsilon_{k} >0$ such that \whp\ the
set of satisfying assignments of $F_{k}(n,rn)$ consists of
$2^{\epsilon_{k} n}$ non--empty cluster-regions, such that
\begin{enumerate}
\item
The diameter of each cluster-region is at most $\alpha_{k}n$.
\item 
The distance between every pair of cluster-regions is at least
$\beta_{k}n$.
\end{enumerate}
\end{theorem}
 
In other words, for all $k \geq 8$, at some point below the
satisfiability threshold, the set of satisfying assignments consists
of exponentially many, well-separated cluster-regions.  The picture
suggested by Theorem~\ref{basic} comes in sharper focus for large
$k$. In particular, for sufficiently large $k$, sufficiently close to
the threshold, the cluster regions become arbitrarily small and
maximally far apart (while remaining exponentially many). The
following result gives a quantitative version of this fact.
\begin{theorem}\label{sharp_basic}
For any $0 < \delta <1/3$, if $r = (1 - \delta) 2^{k} \ln 2$, then for
all $k \ge k_{0}(\delta)$, Theorem~\ref{basic} holds with
\begin{eqnarray*}
  \alpha_{k} 	= \frac{1}{k} \enspace , \qquad
  \beta_{k} 	= \frac{1}{2}-\frac{5}{6}\sqrt{\delta}
  \enspace , \qquad
  \epsilon_{k} 	= \frac{\delta}{2} - 3k^{-2} \enspace .
\end{eqnarray*}
\end{theorem}
It is worth noting that, as we will show shortly,
\begin{remark}
Theorems~\ref{basic} and~\ref{sharp_basic} remain valid for any
definition of clusters in which a pair of assignments are deemed
adjacent whenever their distance is at most $f(n)$ for some function
$f(n)=o(n)$.
\end{remark}
 
Our main result in this paper comes from ``looking inside" clusters
and proving the existence of variables which take the same value in
all the truth assignments of a cluster.  More formally,
\begin{definition}
The \textbf{projection} of a variable $x_i$ over a set of satisfying
assignments $C$, denoted as $\pi_i(C)$, is the union of the values
taken by $x_i$ over the assignments in $C$. If $\pi_i(C) \neq \{0,1\}$
we say that $x_i$ is \textbf{frozen} in $C$.
\end{definition}

No previous results were known about the existence of frozen
variables. The existence of such variables is a fundamental
underpinning of the approximations implicit in the Survey Propagation
algorithm. To prove that random formulas have frozen variables we, in
fact, prove that random formulas have non-trivial cores for all $k
\geq 9$, thus also answering the main question of~\cite{elitza} (we
postpone the definition of cores until Section~\ref{sec:frozeness}). A
strength of our approach is that it allows us to prove not just the
existence, but the pervasiveness of frozen variables. Specifically,
Theorem~\ref{gen_c} below asserts that for sufficiently large $k$, as
we approach the satisfiability threshold, the fraction of frozen
variables in every single cluster gets arbitrarily close to 1.
\begin{theorem}[{\bf Main Result}]\label{gen_c}
For every $\parta>0$ and all $k \ge k_0(\parta)$, there exists
$c_k^{\parta} < r_k$, such that for all $r \geq c_k^{\parta}$, \whp\
{\bf every} cluster of $F_k(n,rn)$ has at least $(1-\parta)n$ frozen
variables. As $k$ grows,
\[
\frac{c_k^{\parta}}{2^k \ln 2} \to \frac{1}{1+\parta(1-\parta)}
\enspace .
\]
\end{theorem}
By taking $\alpha=1/2$ in Theorem~\ref{gen_c} we see that for
sufficiently large $k$, every cluster already has a majority of frozen
variables at $r = (4/5 + \delta_k) 2^k \ln 2$, with $\delta_k \to 0$,
\ie for a constant fraction of the satisfiable regime. More generally,
Theorem~\ref{gen_c} asserts that as $k$ grows and the density
approaches the threshold, clusters shrink in volume and grow further
apart by having smaller and smaller internal entropy (more frozen
variables).

The analysis that establishes Theorem~\ref{gen_c} also allows us to
show
\begin{corollary}\label{c_9}
For every $k \geq 9$, there exists $r < r_k$ such that \whp\ {\bf
every} cluster of $F_k(n,rn)$ has frozen variables.
\end{corollary}
It remains open whether frozen variables exist for $k \leq 8$. As we
mentioned above, \cite{elitza} reported experimental evidence
suggesting that frozen variables do {\em not\/} exist for $k=3$. We
will see that our proof also gives evidence in this direction for
small values of $k$.

\section{Clustering: Proof Sketch and Related Work}\label{sec:clustering}

There are two main ingredients for proving Theorems~\ref{basic}
and~\ref{sharp_basic}. The first one excludes the possibility of pairs
of truth assignments at certain Hamming distances.

\subsection{Forbidden distances and their implications for clustering}

\begin{figure}[ht]
\begin{center}
\includegraphics[width=0.5\textwidth]{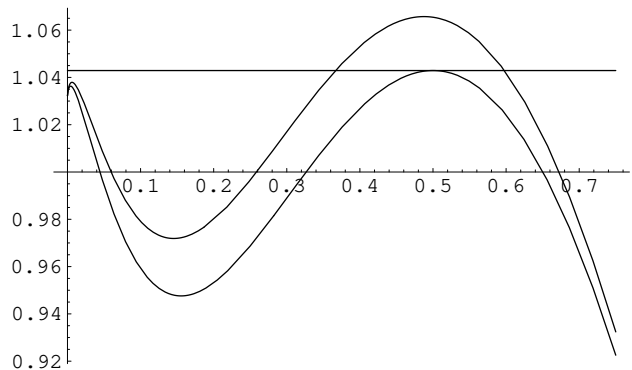}
\end{center}
\caption{Upper curve $\Lambda(\alpha,8,169)$ and lower curve
$\Lambda_b(\alpha,8,169)$ for $\alpha \in [0,3/4]$.}
\label{ploo}
\end{figure}

It is easy to show, see e.g.,~\cite{naesat}, that the expected number
of pairs of satisfying assignments in $F_{k}(n,rn)$ with Hamming
distance $z$ is at most $\Lambda(z/n,k,r)^n$, where
\[
\Lambda(\alpha,k,r) = \frac{2(1-2^{1-k}+2^{-k}
(1-\alpha)^k)^r}{\alpha^{\alpha} (1-\alpha)^{1-\alpha}}  \enspace .
\]
Therefore, if for some $k,r$ and $z=\alpha n$ we have
$\Lambda(\alpha,k,r)<1$, it immediately follows by the union bound
that \whp\ in $F_k(n,rn)$ no pair of satisfying assignments has
distance $z$.  This observation was first made and used
in~\cite{phys_clus_j}.  In Figure~\ref{ploo} we draw the function
$\Lambda$ (upper curve), and a related function $\Lambda_{b}$ (lower
curve, to be discussed shortly), for $\alpha \in [0,3/4]$ with $k=8$
and $r=169$. Recall that, by the results in~\cite{AP}, $F_8(n, 169 n)$
is \whp\ satisfiable and, thus, excluding the possibility of
satisfying pairs at certain distances is a non-vacuous statement. We
see that $\Lambda(\alpha,8,169)<1$ for $\alpha \in [0.06,0.26] \cup
[0.68,1]$, implying that \whp\ in $F_8(n, 169 n)$ there is no pair of
satisfying assignments with Hamming distance $\alpha n$ for such
values of $\alpha$.\smallskip
  
Establishing that there exists a distance $z$ such that there are no
pairs of assignments at distance $z$ immediately implies an upper
bound on the diameter of every cluster. This is because if a cluster
$C$ has diameter $d$, then it must contain pairs of solutions at every
distance $1 \leq t \leq d$. To see this, take any pair
$\sigma_1,\sigma_2 \in C$ that have distance $d$, any path from
$\sigma_1$ to $\sigma_2$ in $C$, and observe that the sequence of
distances from $\sigma_1$ along the vertices of the path must contain
every integer in $\{1,\ldots,d\}$. Therefore, if $Makis = \Makis_{k,r}
\equiv \inf \{\alpha: \Lambda(\alpha,k,r) <1\}$, then \whp\ every
cluster in $F_k(n,rn)$ has diameter at most $\Makis n$.

If we can further prove that $\Lambda(\alpha,k,r) < 1$ in an interval
$(\a,\beta)$, then we can immediately partition the set of satisfying
assignments into well-separated regions, as follows. Start with any
satisfying assignment $\sigma$, let $C$ be its cluster, and consider
the set $R(C) \subseteq \{0,1\}^n$ of truth assignments that have
distance at most $\a n$ from $C$ and the set $B(C) \subseteq
\{0,1\}^n$ of truth assignments that have distance at most $\beta n$
from $R(C)$. Observe now that the set $B(C) \setminus R(C)$ cannot
contain any satisfying truth assignments, as any such assignment would
be at distance $\a n < d < \beta n$ from some assignment in $C$. Thus,
the set of satisfying assignments in $R(C)$ is a union of clusters
(cluster-region), all of which have distance at least $\beta n$ from
any cluster not in the region. Repeating this process until all
satisfying assignments have been assigned to a cluster region gives us
exactly the subsets of Theorems~\ref{basic}
and~\ref{sharp_basic}. Moreover, note that this arguments bounds the
diameter of each entire cluster-region, not only of each cluster, by
$\a n$.
\begin{remark}
The arguments above remains valid even if assignments are deemed
adjacent whenever their distance is bounded by $f(n)$, for any
$f(n)=o(n)$. As a result, Theorems~\ref{basic} and~\ref{sharp_basic}
remain valid as stated for any definition of clusters in which
assignments are deemed to belong in the same cluster if their distance
is $o(n)$.
\end{remark}

\subsection{Establishing exponentially many clusters}

Proving the existence of exponentially many non-empty cluster regions
requires greater sophistication and leverages in a strong way the
results of~\cite{AP}. This is because having $\Lambda(\alpha,k,r)>1$
for some $\alpha,k,r$ does {\em not\/} imply that pairs of satisfying
assignments exist for such $\alpha,k,r$: in principle, the behavior of
$\Lambda$ could be determined by a tiny minority of solution-rich
formulas. Hence the need for the second moment
method~\cite{naesat,AP}. Specifically, say that a satisfying
assignment is {\em balanced\/} if its number of satisfied literal
occurrences is in the range $km/2 \pm \sqrt{n}$, and let $X$ be the
number of balanced assignments in $F_k(n,rn)$. In~\cite{AP}, it was
shown that $\ex[X]^2 = \Lambda_b(1/2,k,r)^n$ and
\[
\ex[X^2] < C\times \max_{\alpha \in [0,1]}\Lambda_b(\alpha,k,r)^n
\enspace ,
\]
for some explicit function $\Lambda_b$ and constant $C=C(k)>0$.  It
was also shown that for all $r < 2^k \ln 2 - k$, the maximum of
$\Lambda_b$ occurs at $\alpha=1/2$, implying that for such $k,r$ we
have $\ex[X^2] < C\times \ex[X]^2$. By the Payley-Zigmund inequality,
this last fact implies that for any $t \leq \ex[X]$,
\begin{equation}\label{eq:pz}
  \Pr[X > t] \geq \frac{(\ex[X]-t)^2}{\ex[X^2]} \enspace .
\end{equation}

In~\cite{AP}, inequality~\eqref{eq:pz} was applied with $t=0$, per the
``second moment method", establishing that for $r < 2^k \ln 2 - k$,
$F_k(n,rn)$ has at least one (balanced) satisfying assignment with
probability at least $1/C$. Taking $t = \ex[X]/{\mathrm{poly}}(n)$,
implies that $X$ is within a polynomial factor of its expectation
$\Lambda_b(1/2,k,r)^{n/2}$, also with constant probability. Since the
property ``has more than $q$ satisfying assignments" has a sharp
threshold~\cite{fried}, this assertion implies that for every $r < 2^k
\ln 2 -k$, $F_k(n,rn)$ has at least $\Lambda_b(1/2,k,r)^{n/2}
/{\mathrm{poly}}(n)$ satisfying assignments \whp

To prove that there are exponentially many clusters, we divide the
above lower bound for the total number of satisfying assignments with
the following upper bound for the number of truth assignments in each
cluster-region. Recall that $\Makis=\Makis_{k,r} \equiv \inf \{\alpha:
\Lambda(\alpha,k,r) <1\}$ and let
\[
g(k,r) = \max_{\alpha \in [0,\Makis]} \Lambda(\alpha, k,r) \enspace .
\]

If $B$ is the expected number of pairs of truth assignments with
distance at most $\Makis n$ in $F_k(n,rn)$, it follows that $B <
{\mathrm{poly}}(n) \times g(k,r)^n$, since the expected number of
pairs at each distance is at most $\Lambda(\alpha, k,r)^n$ and there
are no more than $n+1$ possible distances. By Markov's inequality,
this implies that \whp\ the number of pairs of truth assignments in
$F_k(n,rn)$ that have distance at most $\Makis n$ is
${\mathrm{poly}}(n) \times g(k,r)^n$. Recall now that \whp\ every
cluster-region in $F_k(n,rn)$ has diameter at most $\Makis
n$. Therefore, \whp\ the total number of pairs of truth assignments in
each cluster-region is at most ${\mathrm{poly}}(n) \times
g(k,r)^n$. Thus, if $g(k,r) < \Lambda_b(1/2,k,r)$, we can conclude
that $F_k(n,rn)$ has at least
\[
1/{\mathrm{poly}}(n) \times
\left(\frac{\Lambda_b(1/2,k,r)}{g(k,r)}\right)^{n/2}
\]
cluster-regions. Indeed, the higher of the two horizontal lines in
Figure~\ref{ploo} highlights that $g(8,169) < \Lambda_b(1/2,8,169)$.

From the discussions in this section we see that to establish
Theorem~\ref{basic} it suffices to prove the following.
\begin{theorem}\label{thm:est}
For every $k \geq 8$, there exists a value of $r<r_{k}$ and constants
$\alpha_{k} < \beta_{k}<1/2$ and $\epsilon_{k} >0$ such that
$\Lambda(\alpha,k,r)<1$ for all $\alpha \in (\alpha_k,\beta_k)$ and
$$
\log_2\left[\left(\frac{\Lambda_b(1/2,k,r)}{g(k,r)}\right)^{1/2}\right]
> \epsilon_k \enspace .
$$
In particular, for any $0 < \delta <1/3$ and all $k \ge
k_{0}(\delta)$, if $r = (1 - \delta) 2^{k} \ln 2$, we can take
\begin{equation}\label{largek}
  \alpha_{k} 	=  \frac{1}{k}  \enspace ,\qquad
  \beta_{k} 	=  \frac{1}{2}-\frac{5}{6}\sqrt{\delta}
  \enspace , \qquad
  \epsilon_{k} 	=  \frac{\delta}{2} -3k^{-2} \enspace .
\end{equation} 
\end{theorem}
Specifically, in Section~\ref{sec:ab} we will prove the claims in
Theorem~\ref{thm:est} regarding $\alpha_k,\beta_k$, while in
Section~\ref{sec:e} we prove the claims regarding $\epsilon_k$.

\subsection{Related Work}\label{sec:related}

The observation that if $\Lambda(\alpha,k,r)<1$ then \whp\ $F_k(n,rn)$
has no pairs of satisfying assignments at distance $\alpha n$ was
first made and used in~\cite{phys_clus_j}. Moreover,
in~\cite{phys_clus_l} the authors gave an expression
$\Lambda_{\ell}(\alpha,k,r)$ for the expected number of {\em locally
maximal\/} pairs of satisfying assignments at each distance, where a
pair $\sigma,\tau$ is locally maximal if there is no variable which
has value 0 in $\sigma$ and flipping its value in both $\sigma$ and
$\tau$ yields a new pair of satisfying assignments. (If a formula has
a pair of satisfying assignments at distance $d$, then it always has a
locally maximal pair at distance $d$). Clearly,
$\Lambda_{\ell}(\alpha,k,r)< \Lambda(\alpha,k,r)$ always, but for
large $k$ and $r = \Theta(2^k$) the difference is minuscule for all
$\alpha$.

The connection between $\Lambda(\a,k,r) <1$ in an interval and
``clustering" was also first made in~\cite{phys_clus_j}.
Unfortunately, in~\cite{phys_clus_j} no concrete definition of
clusters was given and, certainly, no scheme for grouping clusters
into well-separated cluster regions (clusters need not be well
separated themselves). Besides these simple clarifications, our minor
contribution regarding clustering lies in giving rigorous bounds on
the diameter and distance of the cluster regions. The novel one, as we
discuss below, lies in establishing the existence of exponentially
many cluster regions.

Additionally, in~\cite{phys_clus_j,phys_clus_l} the authors derive an
expression for the second moment of the number of {\em pairs of \/}
balanced assignments at distance $\alpha n$, for each $\alpha \in
[0,1]$. Whenever, for some $\alpha,k,r$, the dominant contribution to
this second moment comes from uncorrelated pairs of pairs of balanced
assignments, this implies that with {\em constant} probability
$F_k(n,rn)$ contains at least one (balanced) pair of assignments at
distance $\alpha n$. We note that determining the dominant
contribution to the above second moment rigorously, given
$\alpha,k,r$, is a highly non-trivial problem which the authors tackle
numerically for small $k$, and heuristically for general $k$, \ie they
make a guess for the locus of the maximizer. In particular, this
``fourth moment" optimization problem is {\em much\/} harder than the
already complicated second moment analysis of~\cite{AP}.

Finally, the authors prove that the property ``has a pair of
satisfying assignments at distance $q$" has a sharp threshold, thus
boosting their constant probability result for having a pair of
satisfying assignments at a given distance to a high probability
one. To the best of our understanding, these three are the only
results established in~\cite{phys_clus_l}. Combined, they imply that
for every $k \geq 8$, there is $r<r_k$ and constants
$\alpha_k<\beta_k<c_k<1/2<d_k$, such that in $F_k(n,rn)$:
\begin{itemize}
\item
  W.h.p.\ every pair of satisfying assignments has distance either
  less than $\alpha_{k} n$ or more than $\beta_{k} n$.
\item
  For every $d \in [c_k,d_k] \cdot n$, \whp\ there is a pair of truth
  assignments that have distance $d$.
\end{itemize}
We note that even if the maximizer in the second moment computation
was determined rigorously and coincided with the heuristic guess
of~\cite{phys_clus_l}, the strongest statement that can be inferred
from the above two assertions in terms of ``establishing clustering"
is: for every $k \geq 8$, there is $r < r_k$, such that \whp\
${\mathcal S}(F_k(n,rn))$ has at least two clusters.

In contrast, our Theorem~\ref{basic} establishes that \whp\ ${\mathcal
S}(F_k(n,rn))$ consists of {\em exponentially\/} many, well-separated
cluster regions, each region containing at least one
cluster. Additionally, Theorem~\ref{sharp_basic} establishes that as
$k$ grows and $r$ approaches the threshold, these regions grow
maximally far apart and their diameter vanishes.

\section{Frozen Variables: Survey Propagation and Related Work}
\label{sec:frozeness}

For a cluster $C$, the string $\pi(C)= \pi_1(C),\pi_2(C),\ldots,
\pi_n(C)$ is the {\bf projection} of $C$ and we will use the
convention $\{0,1\} \equiv \ast$, so that $\pi(C) \in
\{0,1,\ast\}^n$. Imagine for a moment that given a formula $F$ we
could compute the marginal of each variable over the cluster
projections, \ie that for each variable we could compute the fraction
of clusters in which its projection is $0,1$, and $\ast$. Then, as
long as we never assigned $1-x$ to a variable which in every cluster
was frozen to the value $x$, we are guaranteed to find a satisfying
assignment: after each step there is at least one cluster consistent
with our choices so far.

Being able to perform the above marginalization seems quite far
fetched given that even if we are handed a truth assignment $\sigma$
in a cluster $C$, it is not at all clear how to compute $\pi(C)$ in
time less than $|C|$. Survey Propagation (SP) is an attempt to compute
marginals over cluster projections by making a number of
approximations. One fundamental assumption underlying SP is that,
unlike the marginals over truth assignments, the marginals over
cluster projections essentially factorize, \ie if two variables are
far apart in the formula, then their joint distribution over cluster
projections is essentially the product of their cluster projection
marginals. Determining the validity of this assumption remains an
outstanding open problem.

The other fundamental assumption underlying SP is that {\em
approximate\/} cluster projections can be encoded as the solutions of
a CSP whose factor graph can be syntactically derived from the input
formula. Our results are closely related to this second assumption and
establish that, indeed, the approximate cluster projections used in SP
retain a significant amount of information from the cluster
projections. To make this last notion concrete and enhance intuition,
we give below a self-contained, brisk discussion of Survey
Propagation. For the sake of presentation this discussion is
historically inaccurate. We attempt to restore history in
Section~\ref{sec:sp_related}.

As we said above, even if we are given a satisfying assignment
$\sigma$, it is not obvious how to determine the projection of its
cluster $C(\sigma)$. To get around this problem SP sacrifices
information in the following manner.

\begin{definition}
Given a string $x \in \{0,1,\ast\}^n$, a variable $x_i$ is
\textbf{free} in $x$ if in every clause $c$ containing $x_i$ or
$\overline{x}_i$, at least one of the other literals in $c$ is
assigned true or $\ast$.

We will refer to the following as a
\begin{center}
  \textbf{\whitening-step:} if a variable is free, assign it $\ast$.
\end{center}
Given $x,y\in \{0,1,\ast\}^{n}$ say that $x$ is dominated by $y$,
written $x\preceq y$, if for every $i$, either $x_{i}=y_{i}$ or
$y_{i}=\ast$.
\end{definition}
Consider now the following process:
\[
\mbox{start at $\sigma$ and apply \whitening\ until a fixed point is reached.}
\]

\begin{lemma}\label{lem:uni}
For every formula $F$ and truth assignment $\sigma \in {\cal S}(F)$,
there is a unique \whitening\ fixed point $w(\sigma)$. If
$\sigma_1,\sigma_2$ belong to the same cluster $C$, then
$w(\sigma_1)=w(\sigma_2)\succeq \pi(C)$.
\end{lemma}

\begin{proof}
Trivially, applying a \whitening\ step to a string $x$ produces a
string $y$ such that $x \preceq y$. Moreover, if $x_i$ was free in
$x$, then $y_i$ will be free in $y$. As a result, if both $y,z\in
\{0,1,\ast\}^n$ are reachable from $x \in \{0,1,\ast\}^n$ by
\whitening\ steps, so is the string that results by starting at $x$,
concatenating the two sequences of operations and removing all but the
first occurrence of each \whitening\ step. This implies that there is
a unique fixed point $w(x)$ for each $x \in \{0,1,\ast\}^n$ under
\whitening. Observe now that if $\sigma,\sigma' \in {\cal S}(F)$
differ only in the $i$-th coordinate, then the $i$-th variable is free
in both $\sigma,\sigma'$ and \whitening\ it in both yields the same
string $\tau$. By our earlier argument,
$w(\sigma)=w(\tau)=w(\sigma')=w_C$, where $C\subseteq {\cal S}(F)$ is
the cluster containing $\sigma,\sigma'$. Considering all adjacent
pairs in $C$, we see that $w_C \succeq \pi(C)$.
\end{proof}

\begin{definition}\label{core_def}
  The \textbf{core} of a cluster $C$ is the unique \whitening\ fixed
  point of the truth assignments in $C$.
\end{definition}

By Lemma~\ref{lem:uni}, if a variable takes either the value 0 or the
value 1 in the core of a cluster $C$, then it is frozen to that value
in $C$.  To prove Theorem~\ref{gen_c} we prove that the core of every
cluster has many non-$\ast$ variables.

\begin{theorem}\label{gen_w}
For any $\parta >0$, let $k_0(\parta)$ and $c_k^{\parta}$ be as in
Theorem~\ref{gen_c}. If $k \geq k_0$ and $r \geq c_k^{\parta}$, then
\whp\ the \whitening\ fixed point of\/ {\bf every} $\sigma \in
\mathcal{S}(F_k(n,rn))$ contains fewer than $\parta \cdot n$ variables
that take the value $\ast$.
\end{theorem}

To prove Theorem~\ref{gen_w} (which implies Theorem~\ref{gen_c}) we
derive sharp bounds for the large deviations rate function of the
\whitening\ process applied to a fixed satisfying assignment. As a
result, we also prove that in the planted-assignment model the cluster
containing the planted assignment already contains frozen variables at
$r \sim (2^k/k) \ln k$. Also, we will see that our proof gives a
strong hint that for small values of $k$, such as $k=3$, for all
densities in the corresponding satisfiable regime, most satisfying
assignments {\em do} converge to $(\ast,\cdots,\ast)$ upon
\whitening.\medskip

We can think of \whitening\ as an attempt to estimate the projection
of $C(\sigma)$ by starting at $\sigma$ and being somewhat reckless. To
see this, consider a parallel version of \whitening\ in which given $x
\in \{0,1,\ast\}^n$ we coarsen all free variables in it
simultaneously. Clearly, the first round of such a process will only
assign $\ast$ to variables whose projection in $C(\sigma)$ is indeed
$\ast$. Subsequent rounds, though, might not: a variable $v$ is deemed
free, if in every clause containing it there is some other variable
satisfying the clause, {\em or\/} a variable assigned $\ast$. This
second possibility is equivalent to assuming that the $\ast$-variables
in the clauses containing $v$, call them $\Gamma_v$, can take joint
values that allow $v$ to not contribute in the satisfaction of any
clause. In general formulas this is, of course, not a valid
assumption. On the other hand, the belief that in random formulas
there are no long-range correlations {\em among the non-frozen\/}
variables of each cluster makes this is a reasonable statistical
assumption: since the formula is random, the variables in $\Gamma_v$
are probably far apart from one another in the factor graph that
results after removing the clauses containing $v$. Thus, indeed, any
subset of variables of $\Gamma_v$ that do not co-occur in a clause
should be able to take {\em any\/} set of joint values. Our results
can be seen as evidence of the utility of this line of reasoning,
since we prove that for sufficiently large densities, the \whitening\
fixed point of a satisfying assignment is {\em never\/}
$(\ast,\ldots,\ast)$. Indeed, as we approach the satisfiability
threshold, the fraction of frozen variables in it tends to 1.\medskip

Of course, while the core of a cluster $C$ can be easily derived given
some $\sigma \in C$, such a $\sigma$ is still hard to come by. The
last leap of approximation underlying SP is to define a set $Z(F)
\subseteq \{0,1,\ast\}^n$ that includes all cluster cores, yet is such
that membership in $Z(F)$ is ``locally checkable", akin to membership
in ${\mathcal S}(F)$. Specifically,
\begin{definition}
A string $x \in \{0,1,\ast\}^n$ is a \textbf{cover} of a CNF formula
$F$ if: (i) under $x$, every clause in $F$ contains a satisfied
literal or at least two $\ast$, and (ii) every free variable in $x$ is
assigned $\ast$, \ie $x$ is $\ast$--maximal.
\end{definition}
Cores trivially satisfy (ii) as fixed points of \whitening; it is also
easy to see, by induction, that any string that results by applying
\whitening\ steps to a satisfying assignment satisfies (i). Thus, a
core is always a cover. At the same time, checking whether $x \in
\{0,1,\ast\}^n$ satisfies (i) can be done trivially by examining each
clause in isolation. For (ii) it is enough to check that for each
variable $v$ assigned $0$ or $1$ in $x$, there is at least one clause
satisfied by $v$ and dissatisfied by all other variables in it. Again,
this amounts to $n$ simple checks, each check done in isolation by
considering the clauses containing the corresponding variable. The
price we pay for dealing with locally-checkable objects is that the
set of all covers $Z(F)$ can be potentially much bigger than the set
of all cores. For example, $(\ast,\cdots,\ast)$ is always a cover,
even if $F$ is unsatisfiable.

The Survey Propagation algorithm can now be stated as follows.
\begin{itemize}
\item[]
  Repeat until all variables are set:
  \begin{enumerate}
  \item
    Compute the marginals of variables over covers.
  \item
    Select a variable with least mass on $\ast$ and assign it the 0/1
    value on which it puts most mass.
  \item
    Simplify the formula.
\end{enumerate}
\end{itemize}

The computation of marginals over covers in the original
derivation~\cite{sp,sp2} of SP was, in fact, done via a message
passing procedure that runs on the factor graph of the original
formula rather than a factor graph encoding covers (more on this in
Section~\ref{sec:sp_related}). Also, in~\cite{sp,sp2}, if a
configuration is reached in which all variables put (nearly) all their
mass on $\ast$, the loop is stopped and a local search algorithm is
invoked. The idea is that when such a configuration is reached, the
algorithm has ``arrived" at a cluster and finding a solution inside
that cluster is easy since only non-frozen variables remain unset.

\subsection{Related Work}\label{sec:sp_related}

The original presentation of Survey Propagation motivated the
algorithm in terms of a number of physical notions (cavities, magnetic
fields, etc.). Specifically, the algorithm was derived by applying the
``cavity method" within a ``1-step Replica Symmetry Breaking" scheme,
with no reference whatsoever to notions such as cluster projections,
cores, or covers (in fact, even clusters where only specified as the
connected components that result when satisfying assignments at
``finite Hamming distance" are considered adjacent). On the other
hand, a very definitive message-passing procedure was specified on the
factor graph of the original formula and the computer code
accompanying the paper and implementing that procedure worked
spectacularly well. Moreover, a notion foreshadowing cores was
included in the authors' discussion of ``Warning Propagation".

Casting SP as an attempt to compute marginals over cores was done
independently by Braunstein and Zecchina in~\cite{alfredo} and Maneva,
Mossel, and Wainwright in~\cite{elitza}. In particular, in both papers
it is shown that the messages exchanged by SP over the factor graph of
the input formula are the messages implied by the Belief Propagation
formalism~\cite{aji} applied to a factor graph encoding the set of all
covers. The first author and Thorpe~\cite{jeremy} have additionally
shown that for every formula $F$, there is a factor graph $G_F$
encoding the set of $F$'s covers which inherits the cycle structure of
$F$'s factor graph, so that if the latter is locally tree-like so is
$G_F$.

In~\cite{elitza}, the authors give a number of formal correspondences
between SP, Markov random fields and Gibbs sampling and note that a
cover $\sigma \in \{0,1,\ast\}^n$ can also be thought of as partial
truth assignment in which every unsatisfied clause has length at least
2, and in which every variable $v$ assigned $0$ or $1$ has some clause
$c$ for which it is essential in $\sigma$, \ie $v$ satisfies $c$ but
all other variables in $c$ are set opposite to their sign in $c$. This
last view motivates a generalization of SP in which marginals are
computed not only over covers, but over all partial assignments in
which every unsatisfied clause has length at least 2, weighted
exponentially in the number of non-essential 0/1 variables and the
number of $\ast$-variables. One particular motivation for this
generalization is that while SP appears to work very well on random
3-CNF formulas, \cite{elitza} gives experimental evidence that such
formulas do not have non-trivial cores, \ie upon \whitening\ truth
assignments end up as $(\ast,\ldots,\ast)$. This apparent
contradiction is reconciled by attributing the success of SP to the
existence of ``near-core" strings allowed under the proposed
generalization.

While~\cite{elitza} provided a framework for studying SP by connecting
it to concrete mathematical objects such as cores and Markov random
fields, it did not provide results on the actual structure of the
solution space of random $k$-CNF formulas. Indeed, motivated by the
experimental absence of cores for $k=3$, the authors asked whether
random formulas have non-trivial cores for any $k$. Our results,
establish a positive answer to this question for all $k\geq 9$.

\section{The Probabilistic Framework} \label{sec:setup}

Theorem~\ref{gen_c} follows from Theorem~\ref{gen_w} and
Lemma~\ref{lem:uni}. To prove Theorem~\ref{gen_w} we say that a
satisfying assignment $\sigma$ is \mbox{$\parta$-coreless} if its
\whitening\ fixed point $w(\sigma)$ has at least $\parta n$
$\ast$-variables. Let $X$ be the random variable equal to the number
of $\parta$-coreless satisfying assignments in a random $k$-CNF
formula $F_k(n,rn)$. By symmetry,
\begin{eqnarray}
  \ex[X] & = & \sum_{\sigma \in \{0,1,\}^n} \Pr[\sigma \mbox{ is
      $\parta$-coreless $\mid$ $\sigma$ is satisfying}] \cdot \Pr[\sigma
    \mbox{ is satisfying}]\\
  & = & 2^n\cdot \left(1-\frac{1}{2^k}\right)^{rn} \cdot \;
  \Pr[\mbox{$\mathbf{0}$ is $\parta$-coreless $\mid$ $\mathbf{0}$ is
  satisfying}]\enspace.\label{conditional}
\end{eqnarray}

Observe that conditioning on ``$\mathbf{0}$ is satisfying" is exactly
the same as ``planting" the $\mathbf{0}$ solution, and amounts to
selecting the $m=rn$ random clauses in our formula, uniformly and
independently from amongst all clauses having at least one negative
literal. We will see that for every $k \geq 3$, there exists
$t_k^{\parta}$ such that
\begin{equation}
  \Pr[{\mathbf 0} \mbox{ is $\parta$-coreless} \mid {\mathbf 0} \mbox{
      is satisfying} ] =
  \begin{cases}\label{tk}
    1-o(1) & {\mbox{if $r < t_k^{\parta}$ \enspace ,}} \cr
    o(1)   & {\mbox{if $r > t_k^{\parta}$ \enspace .}}
  \end{cases}
\end{equation}
In particular, we will see that $t_k^1 \sim (2^k/k) \ln k$.  We find
it interesting (and speculate that it's not an accident) that all
algorithms that have been analyzed so far work for densities below
$t_{k}^1$. More precisely, all analyzed algorithms set each variable
$v$ by considering only a subset of the not yet satisfied clauses
containing $v$ and succeed for some $r < c\, 2^k/k$, where $c$ depends
on the choice of subset.

To prove $\ex[X]=o(1)$ we will derive a strong upper bound for the
probability in~\eqref{tk} when $r \gg t_k^{\parta}$. Specifically, we
will prove that $\Pr[\mbox{$\mathbf{0}$ is $\parta$-coreless $\mid$
$\mathbf{0}$ is satisfying}] < e^{-f(r)n}$ for a function $f$ such
that for all $r \geq c_k^{\parta}$,
\begin{eqnarray}\label{strong}
  2 \cdot \left(1-\frac{1}{2^k}\right)^{r} \cdot e^{-f(r)} < 1 \enspace.
\end{eqnarray}
By~\eqref{conditional}, for all such $r$ we have $\ex[X] =o(1)$
and Theorem~\ref{gen_w} follows.

\subsection{\Whitening\ as Hypergraph Stripping}

Given any CNF formula $F$ and any $\sigma \in {\mathcal S}(F)$ it is
easy to see that $w(\sigma)$ is completely determined by the set of
clauses $U(\sigma)$ that have precisely one satisfied literal under
$\sigma$. This is because after any sequence of \whitening \ steps
applied to $\sigma$, a clause that had two or more satisfied literals
under $\sigma$, will have at least one satisfied literal or at least
two $\ast$ and thus never prevent any variable from being
free. Therefore, to coarsen a truth assignment $\sigma$ it is enough
to consider the clauses in $U(\sigma)$. Let us say that a variable $v$
is unfrozen if there is no clause in which it is the unique satisfying
variable and let us say that a clause is unfrozen if it contains an
unfrozen variable. It is now easy to see that \whitening\ $\sigma$ is
equivalent to starting with $U$ and removing unfrozen clauses, one by
one, in an arbitrary order until a fixed point is reached, \ie no
unfrozen clauses remain. Variables occurring in any remaining (frozen)
clauses are, thus, frozen in $w(\sigma)$ (to their value in $\sigma$),
while all other variables are assigned $\ast$. This view of
\whitening\ as repeated removal of clauses from $U(\sigma)$ will be
very useful in our probabilistic analysis below.

To estimate $\Pr[{\mathbf 0} \mbox{ is $\parta$-coreless} \mid
{\mathbf 0} \mbox{ is satisfying} ]$ we consider a random $k$-CNF
formula with $rn$ clauses chosen uniformly among those satisfying
$\mathbf{0}$. To determine $w(\mathbf{0})$, by our discussion above,
it suffices to consider the clauses in our formula that have precisely
one satisfied (negative) literal. The number of such clauses is
distributed as
\[
m = \mathrm{Bin}\left(rn,\frac{k}{2^k-1}\right) \enspace .
\]

It will be convenient to work in a model where each of these $m$
clauses is formed by choosing 1 negative literal and $k-1$ positive
literals, uniformly, independently {\em and with replacement}. (Since
$m = O(n)$, by standard arguments, our results then apply when
replacement is not allowed and the original number of clauses is $rn -
o(n)$.) We think of the $k$ literals in each clause as $k$ balls; we
paint the single satisfied literal of each clause red, and the $k-1$
unsatisfied literals blue. We also have one bin for each of the $n$
variables and we place each literal in the bin of its underlying
variable. We will use the term ``blue bin" to refer to a bin that has
at least one blue ball and no red balls. With this picture in mind, we
see that the $\ast$-variables in $w(\mathbf{0})$ correspond precisely
to the set of empty bins when the following process terminates:

\begin{enumerate}
\item\label{qlegal}
  Let $v$ be any blue bin; if none exists exit.
  \hfill{\em{\small \%Identify an unfrozen variable $v$ if one
  exists.}}
\item
  Remove any ball from $v$.
  \hfill{\em{\small \%Remove the occurrence of $v$ in some
  (unfrozen) clause $c$.}} 
\item\label{random} 
  Remove $k-2$ random blue balls.
  \hfill{\em{\small \%Remove the other $k-2$ unsatisfied
  literals of $c$.}} 
\item\label{or:red}
  Remove a random red ball.
  \hfill{\em{\small \%Remove the satisfied literal in $c$.}}
\end{enumerate} 
Note that the above process removes exactly one clause (1 red ball and
$k-1$ blue balls) in each step and, therefore, if we pass the
condition in Step 1, there are always suitable balls to remove. To
give a lower bound on the probability that the process exits before
$m$ steps (thus, reaching a non-trivial fixed point), we will give a
lower bound on the probability that it exits within the first $i =
\alpha m$ steps, for some carefully chosen $\alpha=\alpha(k,r) \in
(0,1)$. In particular, observe that for the process to not exit within
the first $i$ steps it must be that:
\begin{eqnarray}
  \mbox{At the beginning of each of the first $i$ steps there is at
  least one blue bin.} \label{master_prob}
\end{eqnarray}

To bound the probability of the event in~\eqref{master_prob} we will
bound the probability it occurs in the following simplified
process. The point is that this modified process is significantly
easier to analyze, while the event in~\eqref{master_prob} is only
(slightly) more likely (for the values of $k,r$ of interest to us).
\begin{enumerate}
\item[(a)]
  Let $v$ be any blue bin; if none exists go to Step~(c).
\item[(b)]
  Remove any ball from $v$.
\item[(c)]
  Remove a random red ball.
\end{enumerate}

\begin{lemma}\label{ui}
  The event in~\eqref{master_prob} is no less likely in the modified
  process than in the original process.
\end{lemma}

We prove Lemma~\ref{ui} below. To bound the probability of the event
in~\eqref{master_prob} in the modified process we argue as
follows. Let $q$ be the number of bins which do not contain any red
ball after $i$ steps and let $b$ be the original number of blue balls
in these $q$ bins. If $b < i$, then after $b$ steps of the modified
process every non-empty bin will contain at least one red ball, since
up to that point we remove precisely one blue ball per
step. Therefore, the probability of the event in~\eqref{master_prob}
is bounded above by the probability that $b \geq i$. To bound this
last probability we observe that the red balls in the modified process
evolve completely independently of the blue balls. Moreover, since we
remove exactly one red ball in each step, the state of the red balls
after $i$ steps is distributed exactly as if we had simply thrown
$m-i$ red balls into the $n$ bins.

So, all in all, given a random $k$-CNF formula $F$ with $rn$ clauses
and a fixed $0 \le i \le n$, conditional on $\mathbf{0}$ satisfying
$F$, the probability that the \whitening\ process started at
$\mathbf{0}$ fails to reach a fixed point within $i$ steps is bounded
by the probability that $b \geq i$, where
\begin{eqnarray}
  b & = & \mathrm{Bin}\left((k-1)m,\frac{q}{n}\right) \enspace ,
  \quad\mbox{where} \label{b_dist}  \\ 
  m & = & \mathrm{Bin}\left(rn,\frac{k}{2^k-1}\right) \enspace ,
  \quad\mbox{and} \label{ss_dist} \\ 
  q & = & \mathrm{Occ}\left(m-i,n\right) \enspace , \label{q_dist}
\end{eqnarray}
where $\mathrm{Occ}(x,y)$ is the distribution of the number of empty
bins when we throw $x$ balls into $y$ bins.

As a result, given $k,r$, our goal is to determine a value for $i$
that minimizes $\Pr[b \geq i]$. Before we delve into the probabilistic
calculations, in the next section we comment on how our analysis
relates to the planted assignment problem and to the existence of
non-trivial cores for small values of $k$.

\begin{proof}[Proof of Lemma~\ref{ui}]
Consider a process which is identical to the original process except
with Step~\ref{random} removed. We will call this the intermediate
process. We begin by proving that the original and the intermediate
processes can be coupled so that whenever the event
in~\eqref{master_prob} occurs in the original process it also occurs
in the intermediate process.

First, observe that the evolution of the red balls in both processes
is purely random and therefore can be assumed to be identical, \ie we
can think of the original process as making a genuine random choice in
Step~\ref{or:red} and the intermediate process as mimicking that
choice. (We think of all balls as carrying a distinct identifier.)
Similarly, we can assume that originally, the placement of the blue
balls in bins is identical for the two processes.

Let us say that a pair of blue ball placements is good if in every bin
the set of blue balls in the original process is a subset of the set
of blue balls in the intermediate process. Clearly, whenever we are in
a good configuration, since the placement of the red balls is
identical in the two processes, any choice of bin and ball of the
original process in Steps 1,2, is an available choice for the
intermediate process. Moreover, if the intermediate process mimics
these choices, this results is a new good pair of blue ball
placements. By induction, since the original pair of blue ball
placements is good, if the event in~\eqref{master_prob} occurs in the
original process it also occurs in the intermediate process.

Next, we compare the intermediate process to the modified process
observing that they are identical except that in the event that we run
out of bins containing only blue balls the intermediate process stops,
while the modified process carries on. Therefore, we couple the two as
follows: the modified process mimics the intermediate process for as
long as the event in~\eqref{master_prob} does not occur, and makes its
own random choices afterwards. Therefore, if the event
in~\eqref{master_prob} occurs in the intermediate process it also
occurs in the modified process.
\end{proof}

\section{The planted assignment model and small values of $k$}
\label{sec:plantedModel}

Conditional on $\mathbf{0}$ being satisfying, analyzing
$w(\mathbf{0})$ is exactly the same as working in the ``planted
assignment model" and analyzing the core of the cluster containing the
planted assignment. This is rather easy to do if we are content with
results holding with probability $1-o(1)$. Specifically,
by~\eqref{b_dist},\eqref{ss_dist},\eqref{q_dist} and standard
concentration results it follows immediately that if $i = \alpha m$
then \whp
\begin{eqnarray}
m = \lambda \cdot n + o(n),&&\mbox{ where } \lambda=\frac{rk}{2^k-1}\\
q = \gamma \cdot n + o(n),&& \mbox{ where } \gamma=
\exp\left(-\lambda(1-\alpha)\right)\\
b = \beta \cdot n + o(n),&&\mbox{ where } \beta = (k-1)\gamma \lambda
\enspace .
\end{eqnarray}

With these conditionals in place, we can next determine the mean path
of the \whitening\ process using the method of differential
equations~\cite{worm}, \ie the number of red and blue balls after each
step, up to $o(n)$. In particular, this allows us to show that
\begin{claim}
For every $k \geq 3$, there exists a critical value $t_k^1$ such that
if $r<t_k^1$ then \whp\ $w(\mathbf{0})=(\ast,\ldots,\ast)$, while if
$r>t_k^1$ then \whp\ a bounded fraction of the variables in
$w(\mathbf{0})$, and therefore in $C(\mathbf{0})$, are frozen.
\end{claim}

In the table below we give the value of $t_k^1$ for some small values
of $k$ (rounding to two decimals). By lower/upper below we mean the
best known lower/upper bound for satisfiability threshold.
\begin{table}[ht]\label{tab:valu}
\centering
\[
\begin{array}{c|ccccccc}
k             &  3   & 4     & 5     & 6     & 7     \\ \hline
\mbox{Lower}  & 3.52 & 7.91  & 18.79 & 40.62 & 84.82 \\
\mbox{Upper}  & 4.51 & 10.23 & 21.33 & 43.51 & 87.88 \\
\mbox{$t_k^1$}& 5.72 & 11.58 & 21.75 & 40.13 & 73.88 \\
\mbox{$u_k$}  & 6.25 & 12.34 & 22.90 & 41.95 & 76.84 \\
\end{array}
\]
\label{marianthi} 
\end{table}

We see that for $k=3,4,5$, the probability that $\mathbf{0}$ has a
non-trivial \whitening\ fixed point conditional on being satisfying,
tends to 0 for all densities in the satisfiable regime. Clearly,
conditioning on ``$\mathbf{0}$ is satisfying", is not the same as
picking a ``typical" satisfying assignment. Nevertheless, the gap
between $t_k^1$ and the best threshold upper bound for $k=3$ is
sufficiently large to strongly suggest that below the satisfiability
threshold, most satisfying assignments do arrive at
$(\ast,\ldots,\ast)$ upon \whitening. This is consistent with the
experimental results of~\cite{elitza}, who first raised this
possibility.  That said, a distinction worth mentioning is that even
if the \whitening\ procedure arrives at $(\ast,\ldots,\ast)$ from
most/all satisfying assignments there can still be (many) frozen
variables: simply, their corresponding clusters may not be be compact
(``cube-like") enough for \whitening\ to discover their core.

We now comment on the couple of simplifications of the original
process that we introduced in the previous section in order to get a
process that is easier to analyze. As we showed, these simplifications
only increase the probability of the event in~\eqref{master_prob} and
it is natural to wonder if this increase is significant, allowing for
the possibility that our analysis can be made much sharper. Below we
give evidence that this is not the case. In particular, if each of
$m,q,b$ can be assumed to be within $o(n)$ of its expected value, then
the inequality $b \geq i$ in the modified process is equivalent to
\begin{eqnarray*}
r < \frac{2^k-1}{k} \cdot
\frac{\ln\left(\frac{k-1}{\alpha}\right)}{1-\alpha} \equiv u_k(\alpha)
\enspace .
\end{eqnarray*}
In the table above we give the value of $u_k
=\min_{\alpha}u_k(\alpha)$ for some small values of $k$. As we can
see, these values are quite close to $t_k^1$ and get relatively closer
as $k$ is increased. In other words, considering the modified process
does not cause too big a loss in the analysis. Indeed, taking e.g.,\
$\alpha=1/\ln k$, already gives $u_k \to (2^k/k) \ln k$, which is
consistent with the physics prediction that $t_k^1 \to (2^k/k) \ln k$.

Of course, if one is interested in establishing that certain
properties of $w(\mathbf{0})$ hold with exponentially small failure
probability, as we do, then conditioning that $m,q,b$ are within
$o(n)$ of their expectation is not an option. One has to do a large
deviation analysis of all these variables and their interactions in
the \whitening\ process and determine the dominant source of
fluctuations. This is precisely what we do with respect to the event
$b \geq i$ in the modified process.

\section{Large Deviations}\label{sec:ana}

It is well-known that if $np>0$ then for every $\delta \geq -1$,
\[
\Pr[\mathrm{Bin}(n,p) = (1+\delta)np] \leq F(np,\delta) \enspace ,
\]
where
\[
F(x,y) = \exp(-x[(1+y)\ln(1+y)-y]) \enspace .
\]
A similar large deviations bound was shown in~\cite{pavlos} for the
number of empty bins in a balls-in-bins experiment (Theorem~3). That
is, for every $\delta \geq -1$,
\[
\Pr[\mathrm{Occ}(m,n) = (1+\delta)e^{-m/n}] \leq F(ne^{-m/n},\delta)
\enspace .
\]

\subsection{Application}

Write $r = \lambda (2^k-1)/k$ and fix $\delta,\epsilon,\zeta \geq -1.$
Write $\rho = \lambda(1+\delta)(1-\alpha)$ in order to compress the
expressions below. The probability that
\begin{eqnarray}
m & = & (1+\delta) \ex[m] = (1+\delta) \frac{rk}{2^k-1} \cdot n =
  (1+\delta) \lambda \cdot n \enspace ,\label{mdev} \\
q & = & (1+\zeta) \ex[q | m] = (1+\zeta)
  \exp\left(-\frac{m-i}{n}\right) \cdot n = (1+\zeta) e^{-\rho}\cdot n
  \enspace , \label{qdev}\\
b & = & (1+\epsilon) \ex[b|q,m] = (1+\epsilon) (k-1)m \cdot
  \frac{q}{n} = (1+\delta)(1+\epsilon)(1+\zeta)\lambda(k-1) 
  e^{-\rho}\cdot n \, , \label{bdev}
\end{eqnarray}
is bounded by
\begin{equation}\label{eq:kota}
F(\lambda n,\delta) \cdot  F(e^{-\rho}n,\zeta) \cdot
F((1+\delta)(1+\zeta)\lambda(k-1)e^{-\rho}n,\epsilon) \enspace .
\end{equation}
We write this as $e^{-n\,\Omega}$,  where
\[
\Omega \equiv \lambda \omega(\delta) + e^{-\rho} \omega(\zeta) +
\lambda(k-1)(1+\delta)(1+\zeta) e^{-\rho} \omega(\epsilon)\;,
\]
with $\omega(x) = (1+x) \ln(1+x) - x$.

Conditional on the events in~\eqref{mdev}--\eqref{bdev} we see
from~\eqref{bdev} that the condition $b\geq i$ becomes $B \geq 0$,
where
\[
B \equiv (1+\epsilon)(1+\zeta)(k-1) e^{-\rho} -\alpha \enspace .
\]
For any fixed $k$, $r$ and $\alpha$ define $\Phi \equiv
\{(\delta,\zeta,\epsilon): B \ge 0\}$. Thus,
\[
\Pr[\mbox{$\mathbf{0}$ is $\parta$-coreless $\mid$ $\mathbf{0}$ is
satisfying}] < \exp(-n \cdot \min_{\Phi}\Omega) \times
\mathrm{poly}(n)
\] 
and to prove that the expected number of $\alpha$-coreless assignments
in $o(1)$, it suffices to prove
\begin{equation}\label{moraki} 
\min_\Phi \Omega >  \ln 2 + r \ln(1-2^{-k})
\equiv s \enspace . 
\end{equation}

\section{Optimization}\label{sec:opt}

To establish~\eqref{moraki} it is enough to prove that the maximum of
$B$ in the variables $\delta$, $\zeta$ and $\epsilon$ under the
condition $\Omega \leq s$ is negative. Considering that the function
$B$ is monotone in the three variables $\delta$, $\zeta$ and
$\epsilon$, the maximizer of $B$ in the region $\Omega \leq s$ has to
be on the boundary, that is for $\Omega=s$. The maximum of $B$ under
the condition $\Omega=s$ corresponds to the extremum of the function
$G=B-\mu(\Omega-s)$, where $\mu$ is a Lagrange multiplier. The
equations for the location of the maximizer are thus given by
derivatives of $G$ with respect to $\delta$, $\zeta$, $\epsilon$ and
$\mu$
\begin{eqnarray}
\partial_\delta G &=& 0 \quad\Rightarrow\quad
\partial_\delta B = \mu\; \partial_\delta \Omega
\label{derdelta}\\
\partial_\zeta G &=& 0 \quad\Rightarrow\quad
\partial_\zeta B = \mu\; \partial_\zeta \Omega
\label{derzeta}\\
\partial_\epsilon G &=& 0 \quad\Rightarrow\quad
\partial_\epsilon B = \mu\; \partial_\epsilon \Omega
\label{dereps}\\
\partial_\mu G &=& 0 \quad\Rightarrow\quad \Omega = s \enspace .
\label{dermu}
\end{eqnarray}

\begin{lemma}\label{extremum}
For any fixed $k$, $r$ and $\alpha\in(0,1)$, at the extremum of the
function $G$ defined by equations (\ref{derdelta})-(\ref{dermu}) the
following assertions hold
\begin{enumerate}
\item $\epsilon$ is non-negative; \item $\zeta$ is non-negative;
\item $\delta$ is non-positive;
\end{enumerate}
\end{lemma}
\begin{proof}
The first assertion follows by observing that $B$ is an increasing
function of $\epsilon$ and $\Omega$ contains $\epsilon$ only in the
third term through $\omega(\epsilon)$.  Therefore, if the maximizer
would be in $\epsilon=\epsilon'<0$, moving to
$\epsilon=\epsilon^{\prime\prime}>0$, with
$\omega(\epsilon')=\omega(\epsilon^{\prime\prime})$, would keep
$\Omega$ constant while increasing $B$.

Combining equations (\ref{derzeta}) and (\ref{dereps}) in order to
remove $\mu$ we have
\[
\partial_\zeta \Omega = \partial_\epsilon \Omega \;
\partial_\zeta B / \partial_\epsilon B
\]
that is
\[
e^{-\rho} \ln(1+\zeta) + (1+\delta) \lambda (k-1) e^{-\rho}
\omega(\epsilon) = (1+\delta) \lambda (k-1) e^{-\rho} (1+\epsilon)
\ln(1+\epsilon)
\]
which, after simplification, reduces to
\begin{equation}\label{eqzeta}
\ln(1+\zeta) = (k-1)\lambda(1+\delta)\epsilon \enspace .
\end{equation}
Thus, for $\delta \geq -1$ and $\epsilon \geq 0$ we have that at the
maximizer $\zeta \geq 0$, proving the second assertion.

Combining equations (\ref{derdelta}) and (\ref{dereps}) we can write
\begin{multline*}
0 = \partial_\delta \Omega - \partial_\epsilon \Omega \;
\partial_\delta B / \partial_\epsilon B =
\lambda e^{-\rho} \Big[ e^{\rho} \ln(1+\delta)
- (1-\alpha) \omega(\zeta) - \rho(1+\zeta)(k-1)\omega(\epsilon) +\\
+ (k-1)(1+\zeta)\omega(\epsilon) +
\rho(1+\epsilon)(1+\zeta)(k-1)\ln(1+\epsilon) \Big]
\end{multline*}
The term within square brackets can be simplified to
\[
e^{\rho} \ln(1+\delta) - (1-\alpha)\omega(\zeta) +
\rho(1+\zeta)(k-1)\epsilon + (k-1)(1+\zeta)\omega(\epsilon)
\]
which, using~\eqref{eqzeta}, implies
\[
e^{\rho} \ln(1+\delta) + (1-\alpha)\zeta +
(k-1)(1+\zeta)\omega(\epsilon)=0 \enspace .
\]
Since for $\zeta,\epsilon \geq 0$ the second and third term terms of
this expression are non-negative, we find that $\delta$ has to be
non-positive at the maximizer in order to satisfy the last equation
(third assertion).
\end{proof}

We next prove some bounds on $\delta$ and $\epsilon$, that hold at the
maximizer.

\begin{lemma}\label{d_easy}
Fix any $r,k$, and $\alpha \in (0,1)$. At the maximizer of $B$,
\[
\delta_0 \equiv -\sqrt{\frac{2s}{\lambda}} \leq \delta \leq 0
\enspace .
\]
\end{lemma}
\begin{proof}
Since $\delta$ in non-positive at the maximizer, we observe that
$\omega(\delta) \geq \delta^2/2$ for $\delta \leq 0$.  Moreover each
of the three terms in $\Omega$ is non-negative for $\delta,\zeta \geq
-1$ and this implies
\[
\lambda\omega(\delta) \leq s \;\Longrightarrow\; \lambda
\frac{\delta^2}{2} \leq s \;\Longrightarrow\; |\delta| \leq
\sqrt{\frac{2s}{\lambda}} \enspace .
\]
\end{proof}

\begin{lemma}\label{e_easy}
Fix any $r,k$, and $\alpha\in(0,1)$. At the maximizer of $B$,
\[
\epsilon < \frac{1-{\alpha}}{k-1} + \frac{\ln 3}
{\lambda(1+\delta_0)(k-1)} \equiv \epsilon_0 \enspace .
\]
\end{lemma}
\begin{proof}
Since every term in $\Omega$ is non-negative, considering the second
term, and using the fact $s = \Omega$ we get
\[
s \ge \omega(\zeta) e^{-\rho} \enspace .
\]
Observe now that for all $x \geq -1$, we have $\omega(x) > x-1$.
Therefore,
\begin{equation}\label{bas}
s e^{\rho} + 2 > 1+\zeta \enspace .
\end{equation}
Using~\eqref{eqzeta} to replace $1+\zeta$ and the fact $s \le \ln 2 <
1$ we can conclude from~\eqref{bas} that
\[
e^{\rho} + 2 > e^{\lambda(1+\delta)(k-1)\epsilon}
\;\Longrightarrow\; \epsilon < \frac{\ln\left(e^{\rho} +
2\right)}{\lambda(1+\delta)(k-1)} \le \frac{\rho + \ln
3}{\lambda(1+\delta)(k-1)} \enspace ,
\]
where for the last inequality we use that
$\rho=\lambda(1+\delta)(1-\alpha)$ is non-negative.  We conclude that
at the maximizer
\[
\epsilon < \frac{1-{\alpha}}{k-1} + \frac{\ln 3}
{\lambda(1+\delta_0)(k-1)} \equiv \epsilon_0 \enspace ,
\]
where $\delta$ has been replaced by its lower bound value.
\end{proof}\medskip

Thus, the stationary point of $G$ must occur in the region $\Lambda =
\{(\delta,\epsilon): \delta_0 \le \delta \le 0, 0 \le \epsilon \le
\epsilon_0\}$. In the next subsections we derive analytical results
for this optimization for all $k\geq 14$, and we summarize results
obtained by numerically finding the stationary point of $G$ for $9 \le
k \le 13$.

\subsection{Proving the existence of frozen variables for $k \geq 14$
  analytically}

For any fixed values of $\delta$ and $\epsilon$, the requirement $B
\geq 0$ implies
\begin{equation}
\zeta \geq \frac{\alpha\;e^{\rho}}{(k-1)(1+\epsilon)} - 1 \enspace
. \label{eq0}
\end{equation}
Plugging this lower bound in the second term of $\Omega$, we see that
the requirement $\Omega = s$ implies
\begin{equation}\label{opala}
\frac{\alpha}{(k-1)(1+\epsilon)}\left[\ln\left(
\frac{\alpha\;e^{\rho}}{(k-1)(1+\epsilon)}\right)-1\right]+
e^{-\rho} \le s \enspace .
\end{equation}
Therefore, it suffices to find $\lambda$ and $\alpha$ such
that~\eqref{opala} cannot be satisfied by any $\delta_0 \le \delta \le
0$ and $0 \le \epsilon \le \epsilon_0$.  This is certainly true if a
lower bound to the l.h.s.\ of~\eqref{opala} makes such an equation
unsatisfied, that is if
\[
\frac{\alpha}{(k-1)(1+\epsilon_0)}\left[\ln\left(
\frac{\alpha\;e^{\lambda(1+\delta_0)(1-\alpha)}}
{(k-1)(1+\epsilon_0)}\right)-1\right] + e^{-\lambda(1-\alpha)}
> s
\]
and the term within the squared brackets above is positive. Thus, to
summarize, it suffices to find $\lambda$ and $\alpha$ such that
\begin{eqnarray}
\frac{\alpha}{(k-1)(1+\epsilon_0)}\left[\ln\left(
\frac{\alpha\;e^{\lambda(1+\delta_0)(1-\alpha)}}
{(k-1)(1+\epsilon_0)}\right)-1\right] > s - e^{-\lambda(1-\alpha)}
> 0 \enspace. \label{second}
\end{eqnarray}

With the change of variable $\lambda = c\,k\,\ln 2$ we have that
\[
s = \ln 2 + c(2^k-1) \ln(1-2^{-k}) \ln 2\le
\left(1-c(1-2^{-k})\right)\ln 2 \;,
\]
and
\[
s-e^{-\lambda(1-\alpha)} \le \left(1-c(1-2^{-k})\right)\ln 2 -
2^{-c k(1-\alpha)} < (1-c) \ln 2 \;,
\]
since for any $c \in [0,1]$, $\alpha \in [0,1]$ and $k > 0$
\[
c \ln 2 < 2^{k(1-c(1-\alpha))} \;.
\]
Therefore, it suffices to establish
\begin{equation}
\frac{\alpha}{(k-1)(1+\epsilon_0)}\left[\ln\left(
\frac{\alpha\;e^{ck(1+\delta_0)(1-\alpha)\ln 2}}
{(k-1)(1+\epsilon_0)}\right)-1\right] \ge (1-c)\ln 2 \label{main}
\enspace .
\end{equation}
Based on Lemmata~\ref{d_easy} and\ref{e_easy} we now introduce simpler
bounds for $\delta$ and $\epsilon$, which hold for all $c \ge 4/5$ and
$k \ge 2$.  Specifically,
\begin{equation}\label{d:sim}
|\delta_0| \le \sqrt{\frac{2(1-c(1-2^{-k}))}{c\,k}} \le
\sqrt{\frac{2(1-3/4\,c)}{c\,k}} \le \frac{1}{\sqrt{k}}\;,
\end{equation}
and
\begin{equation}\label{e:sim}
\epsilon_0 = \frac{1}{k-1} \left(1-\alpha+\frac{\ln 3}
{c\,k(1+\delta_0) \ln 2}\right) \le \frac{2}{k-1}\;.
\end{equation}
Replacing~\eqref{d:sim} and \eqref{e:sim} in~\eqref{main} we have
\[
\frac{\alpha}{k+1} \left[c\,k\,(1-\alpha)
\left(1-\frac{1}{\sqrt{k}}\right) \ln 2 +
\ln\left(\frac{\alpha}{k+1}\right) -1\right] -(1-c)\ln 2 \ge 0 \;.
\]
Solving with respect to $c$, the last inequality becomes
\[
c \ge \frac{1 + \frac{\alpha}{k+1}\left[1 - \ln\left(
\frac{\alpha}{k+1}\right)\right]/\ln 2} {1 + \alpha(1-\alpha)
\frac{1-1/\sqrt{k}}{1+1/k}} \equiv g_c(k,\alpha) \enspace .
\]
For any fixed $\alpha \in (0,1)$, $g_c(k,\alpha)$ is a decreasing
function of $k$, which as $k \to \infty$ tends to
\[
\frac{1}{1+\alpha(1-\alpha)}\enspace.
\]

\begin{figure}[ht]
\begin{center}
\includegraphics[width=0.5\textwidth]{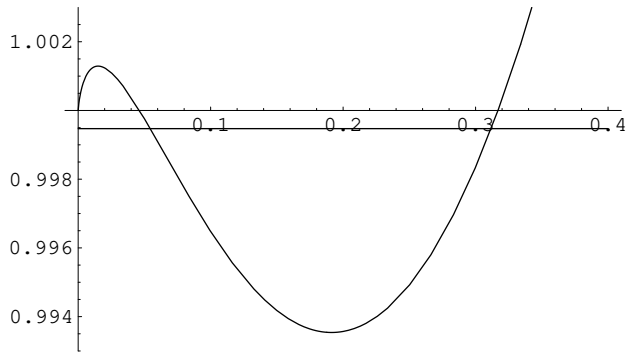}
\end{center}
\caption{The function $g_c(k,\alpha)$ for $k=14$ as a function of
$\alpha$. The horizontal line, slightly below 1, is $\tau_{14} =
0.9994711565304686$.}
\label{k14}
\end{figure}

In order to prove that there exists a choice of $\alpha$ such that
$\min_{\Phi} \Omega > s$ for some $r<r_k$ and all $k \geq
k_0(\alpha)$, we rescale the lower bound for $r_k$ from~\eqref{acp} as
\[
\tau_k \equiv \frac{2^k \ln 2 - \frac{(k+1)\ln 2 +
3}{2}}{(2^k-1)\ln 2} = \frac{1}{1-2^{-k}}-\frac{(k+1)\ln 2
+3}{2(2^k-1)\ln 2} \;,
\]
and observe that $\tau_k$ is increasing in $k$. In Figure~\ref{k14} we
now see that the function $g_c(14,\alpha)$ dips below $\tau_{14}$ for
a certain range of $\alpha$, implying that the left endpoint of this
range is an upper bound on the fraction of unfrozen clauses. For
larger values of $k$ things only get better since $g_c(k,\alpha)$ is
monotonically decreasing with $k$, whereas $\tau_k$ is increasing.
For any fixed value of $\alpha$, $k_0(\alpha)$ can be defined as the
first $k$ value for which $g_c(k,\alpha) < \tau_k$ holds.

\subsection{The $k=9$ case}\label{sec:k9}

Recall that for any fixed $k,r$ and $\alpha$ the function $G$ depends
on four variables: $\delta$, $\epsilon$, $\zeta$ and $\mu$. We will
plot $G$ for $k=9$, $\alpha=0.265$ and a few different values of $r$,
while fixing $\zeta$ and $\mu$ at the value they take at the
stationary point: $\zeta$ is given readily by~\eqref{eqzeta}, and
substituting this value of $\zeta$ into~\eqref{dereps} we get
\[
\mu = \frac{1}{\lambda(1+\delta)\ln(1+\epsilon)} \enspace .
\]

\begin{figure}[ht]
\begin{center}
\includegraphics[width=0.3\textwidth]{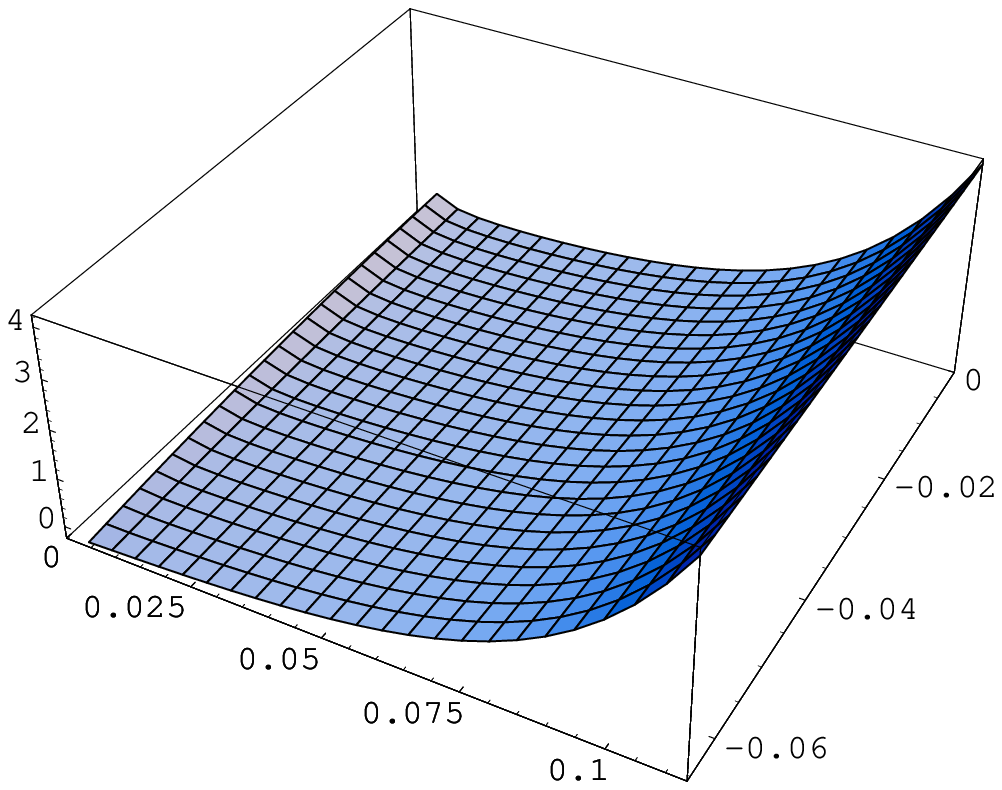}
\includegraphics[width=0.3\textwidth]{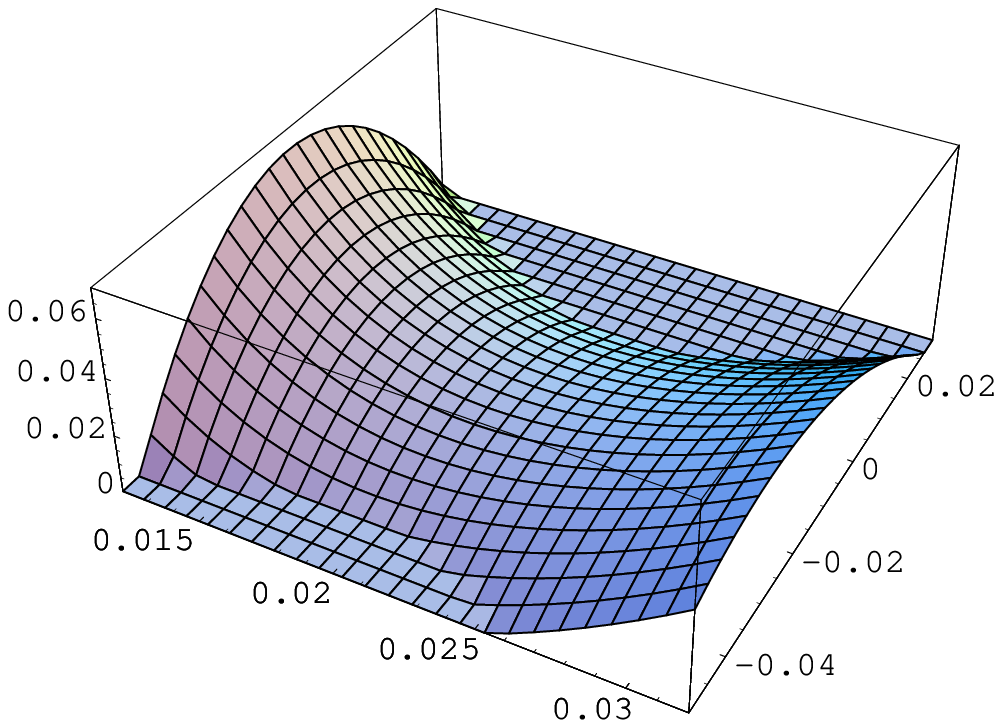}

\includegraphics[width=0.3\textwidth]{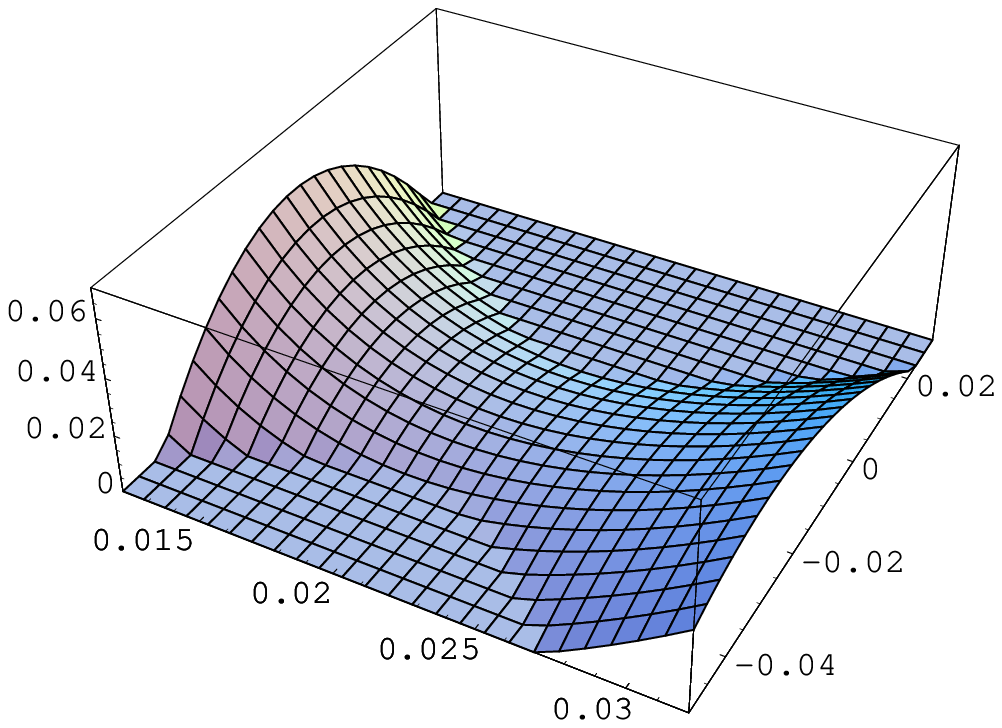}
\includegraphics[width=0.3\textwidth]{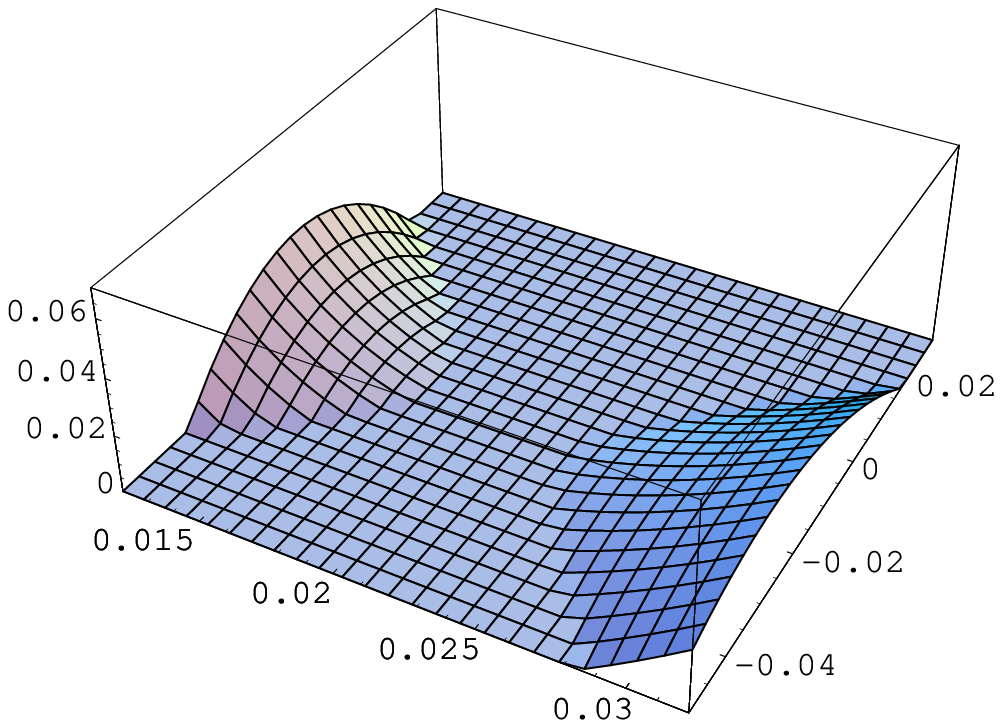}
\end{center}
\caption{$k=9$, $\alpha=0.265$}\label{fig}
\end{figure}

In the upper left panel of Figure~\ref{fig} we show $G$ in the
subregion of $\Lambda$ corresponding to the optimal $\zeta,\mu$ for
$r=347$.  By closer inspection one finds that there is a unique
stationary point in this region. The remaining three plots are zoomed
on the stationary point for $r=347$, $r=347.5$ and $r=348$. It is
clear that the function $G$ at the stationary point is positive for
the first two $r$ values and negative for the third one (for the sake
of clearness, negative values of $G$ are not plotted). Thus, for $k=9$
and $\alpha=0.265$, the critical value of $r$ lies between $347.5$ and
$348$. In the next subsection we determine this critical value
numerically for all $9 \leq k \leq 13$.

\subsection{Optimizing for $9 \le k \le 13$}\label{sec:k9-13}

For any fixed $k$ and $\alpha$ the value of $c_k^{\parta}$, such that
\whp\ clustering exists for $r > c_k^\alpha$, can be computed by
solving numerically~\eqref{derdelta}-\eqref{dermu} together with $G=0$
[which reduces to $B=0$ since $\Omega=s$ by~\eqref{dermu}]. Adding a
sixth equation $\partial_\alpha G = 0$ allows one to minimize
$c_k^{\alpha}$ with respect to $\alpha$ (at some $\alpha_m$) thus
determining the smallest density $c_k$ for which the existence of
frozen variables can be established.  Numerical solutions of these six
equations are given in the table below for $9 \le k \le 13$ along with
the lower bound $r_k$ from~\eqref{acp}.

\begin{center}
\begin{tabular}{|r|r|r|l|l|l|l|l|}
\hline
\multicolumn{1}{|c}{$k$} & \multicolumn{1}{c}{$r_k$} &
\multicolumn{1}{c}{$c_k$} & \multicolumn{1}{c}{$\alpha_m$} &
\multicolumn{1}{c}{$\mu$} & \multicolumn{1}{c}{$\delta$} &
\multicolumn{1}{c}{$\zeta$} & \multicolumn{1}{c|}{$\epsilon$}\\ 
\hline
 9 & 349.92  & 347.84  & 0.265 & 8.037 & -0.015085 & 1.7336 & 0.02083\\
10 & 704.94  & 690.48  & 0.273 & 6.935 & -0.015714 & 2.7134 & 0.02194\\
11 & 1413.90 & 1370.42 & 0.281 & 6.256 & -0.015789 & 4.0330 & 0.02229\\
12 & 2833.12 & 2720.44 & 0.289 & 5.802 & -0.015548 & 5.7977 & 0.02220\\
13 & 5671.90 & 5402.23 & 0.297 & 5.480 & -0.015132 & 8.1457 & 0.02184\\
\hline
\end{tabular}
\end{center}

\section{The existence of cluster regions}\label{sec:ab}

In this section we prove the existence of $\alpha_k,\beta_k$ as in
Theorem~\ref{thm:est}. Let
\begin{eqnarray*}
  h(x) & \equiv & -x\ln x - (1-x) \ln(1-x) \\
  & \le & \ln 2  - 2 (1/2-x)^2 \enspace , \quad \mbox{for any $x \in
  [0,1]$.}
\end{eqnarray*}

We begin by bounding $\ln \Lambda$ from above as follows,
\begin{eqnarray*}
\ln \Lambda(\a, k, \gamma 2^k \ln 2) & = & \ln 2 + h(\a) + \gamma 2^k
\ln 2 \, \ln\left[1 - 2^{1-k} + 2^{-k} (1-\a)^k \right] \\
& < & 2 \ln 2 - 2 \left(1/2 - \a\right)^2 - \gamma \ln 2 \big[2 -
  (1-\a)^k\big] \\
&\equiv & w(\a, k, \gamma)\enspace .
\end{eqnarray*}
We note that for any fixed $k,\gamma$, the function $w(\a,k,\gamma)$
is non-increasing in $k$ and decreasing in $\gamma$. Moreover,
\begin{equation}
\frac{\partial^3 w}{\partial \a^3} = -\gamma \ln 2\, k (k-1) (k-2)
     (1-\a)^{k-3} < 0 \enspace ,
\end{equation}
implying that for any fixed $k,\gamma$, the equation
$w(\alpha,k,\gamma)=0$ can have at most three roots for $\alpha \in
(0,1)$. To bound the location of these roots we observe that for any
$k\ge 8$ and $\gamma > 2/3$,
\begin{eqnarray}
w(0,k,\gamma) &=& (2 - \gamma) \ln 2 - \frac12 > 0 \enspace , \label{magda} \\
w(1/2,k,\gamma) &=& \big[2 - (2 - 2^{-k}) \gamma \big] \ln 2 > 0 \enspace , \\
w(99/100,k,\gamma) &<& w(99/100, 8, 2/3) = -0.0181019... < 0 \enspace,
\label{alkis}
\end{eqnarray}
where the inequality in~\eqref{alkis} relies on the mononicity of $w$
in $k,\gamma$. Therefore, from~\eqref{magda}--\eqref{alkis} we can
conclude that for every $k \ge 8$ and $\gamma > 2/3$, if there exist
$\alpha_k,\beta_k \in (0,1/2)$ such that $w(\alpha_k,k,\gamma) < 0$
and $w(\beta_k,k,\gamma) < 0$, then $\Lambda(\a, k, \gamma 2^k \ln 2)
< 1$ for all $\a \in [\a_k,\beta_k]$. Below we first prove that such
$\alpha_k,\beta_k$ exist for all $k \ge 8$ and then prove that for
sufficiently large $k$, we can take $\alpha_k,\beta_k$ as
in~\eqref{largek}.  \medskip

\begin{itemize}
\item
For $k=8$ it is enough to consider the plot of $\Lambda(\alpha,8,169)$
in Figure~\ref{ploo}. For $k \ge 9$ we take $\g=0.985>2/3$. Note that
$0.985 \cdot2^k \ln 2$ is smaller than the lower bound for $r_k$ given
in~\eqref{acp}, for all $k \ge 9$.
\begin{itemize}
\item
We take $\alpha_k=1/k$. We note that $w(1/9, 9, 0.985) = -0.0451... <
0$ and prove that $w(1/k, k, \g)$ is decreasing in $k$ for any $k \ge
4$ and $\g <1 $ as follows,
\begin{eqnarray*}
\frac{\partial w(1/k, k, \g)}{\partial k} & = & 
\g \ln 2 \left(1 - \frac1k\right)^{k-1} \left[ \frac1k + \left(1 -
  \frac1k\right) \ln\left(1 - \frac1k\right) \right] - \frac{4}{k^2}
\left(\frac12 - \frac1k \right) \\
& < & \g \ln 2 \left(1 - \frac1k\right)^{k-1} \frac{1}{k^2} - \frac{4}{k^2}
\left(\frac12 - \frac1k \right) \\
& < & \frac{1}{k^2} \left(\ln 2\, -2
+\frac4k \right) \\ & <&   0  \enspace .
\end{eqnarray*}
\item
We take $\beta_k = 3/8$. We note that $w(\a, k, \g)$ is non-increasing
in $k$ when $\a$ and $\gamma$ are fixed and that $w(3/8, 9,0.985) =
-0.000520265... < 0$.
\end{itemize}
\item
For the setting where $r=(1-\delta)2^k \ln 2$, we will additionally
use that $-2x\ln 2 <\ln(1-x)<-x$ for all $0<x<1/2$ to establish that
for any $1 \le c <k/2$,
\begin{eqnarray}
  \ln \Lambda(c/k,k,r)	& = & \ln 2 + h(c/k)
  +r\ln(1-2^{1-k}+2^{-k}(1-c/k)^k) \nonumber \\
  & < &  \ln 2 +(c/k) (\ln k+2 \ln 2)-r(2^{1-k}-2^{-k}(1-c/k)^k)
  \label{holi} \enspace .
\end{eqnarray}
Substituting $r = \gamma 2^k \ln 2$ into~\eqref{holi} we get 
\begin{equation}\label{eq:hal}
\ln \Lambda(c/k,k,\gamma 2^k \ln 2) < \ln 2 (1-2\gamma+\gamma e^{-c})
+ (c/k) (\ln k + 2 \ln 2) \enspace .
\end{equation}
\begin{itemize}
\item
If $c=1$ and $\gamma > \frac{1}{2-1/e}=0.612...$, then~\eqref{eq:hal}
implies that $\ln \Lambda(1/k,k,\gamma) <0$ for all sufficiently large
$k$.
\item
If $\g = (1 - \delta) > 2/3$, then for any $1 < \lambda \le 3/(4 \ln
2) = 1.082...$
$$
w(1/2 - \sqrt{\lambda \delta \ln 2}, k, 1-\delta) =
-2 (\lambda-1) \delta \ln 2 + (1-\delta) \ln 2 \left(\frac{1}{2} +
\sqrt{\lambda \delta \ln 2}\right)^k \enspace ,
$$
which is negative for all sufficiently large $k$.
The choice $\beta_k=1/2-(5/6)\sqrt{\delta}$ corresponds to $\lambda=(5/6)^2/\ln 2 = 1.00187...$, which is a valid value. For $k$ large enough we have $\a_k=1/k <\beta_k=1/2-5\sqrt{\delta}/6$ for any $\delta \in (0,1/3)$.
\end{itemize}

\end{itemize}

\section{The existence of exponentially many cluster regions}
\label{sec:e}

We will use the following two lemmata. 
\begin{lemma}\label{lemvol}
If $\gamma \ge 49/50$ and $k>11$, or $\gamma \in (2/3,1)$ and $k>15$,
\begin{equation}
\ln g(k,\gamma 2^k \ln 2) \le (1-\gamma)\ln 2   + \left(1+\frac{9 \ln
  2}{16} \right) k^{-2}  \enspace . 
\end{equation}
\end{lemma} 
\begin{lemma}\label{lembal}
For all $k\ge 8$,  
$$
\ln \Lambda_b(1/2,k,\gamma 2^k \ln 2) \ge 2 \ln 2 [1 -  \gamma m(k)]
\enspace , 
$$ 
where
\begin{multline*}
m(k) = 1 + \frac{2k+3}{2} 2^{-k} + \frac{3k^2+6k-4}{2} 2^{-2k} +
\frac{13k^2-12k+1}{2} 2^{-3k} \\
+ (6k^3-13k^2+2k) 2^{-4k}+ \frac{9k^4-24k^3+10k^2}{2} 2^{-5k} +
(9k^4-6k^3) 2^{-6k} + \frac{9}{2} k^4 2^{-7k} \enspace .
\end{multline*}
\end{lemma}

Combining the two lemmata above we get that if $r = \gamma 2^k \ln 2$
and either $\gamma = 49/50$ and $k>11$, or $\gamma \in (2/3,1)$ and
$k>15$, then
\begin{equation}\label{entrop}
\log_2
\left[\left(\frac{\Lambda_b(1/2,k,r)}{g(k,r)}\right)^{1/2}\right] >
\frac{1}{2 \ln 2} 
\left[\ln 2 (1+\gamma -2 \gamma m(k)) - \left(1+\frac{9 \ln
    2}{16}\right)k^{-2}\right] \enspace ,
\end{equation}
where $m(k)$ is as in Lemma~\ref{lembal}. It is not hard to check that
$m(k)$ is decreasing in $k$.
\begin{itemize}
\item
For $8 \le k \le 12$, the existence of $\epsilon_k>0$ can be verified
by plotting $\Lambda$ and $\Lambda_b$ and noting that
$$
\Lambda_b(1/2,k,r) > \max_{\alpha \in [0,\Makis]} \Lambda(\alpha, k,r)
\enspace , 
$$
both when $k=8$ and $r=169$ and when $9 \leq k \leq 12$ and $r=0.985
\cdot 2^k \ln 2$. For $k>12$ and $\gamma = 0.985$, the existence of
$\epsilon_k >0$ follows from the fact that the expression inside the
square brackets in~\eqref{entrop} is positive when $k=13$ and
$\gamma=0.985$ and $m(k)$ is decreasing in $k$.
\item
For the setting where $r=(1-\delta)2^k \ln 2$, we note that the limit
of the expression inside the square brackets in~\eqref{entrop} as
$k\to \infty$ is $(1-\gamma) / 2$. In particular, writing
$r=(1-\delta) 2^k \ln 2$, it is not hard to show that the right hand
side of~\eqref{entrop} is greater than $\delta/2 - 3/k^2$ for all $k
\ge k_0(\delta)$.
\end{itemize}

\subsection{Proof of Lemma~\ref{lemvol}: The volume of the largest cluster}

Below, we consider $k$ and $r$ to be fixed, so that all derivatives
are with respect to $\alpha$. Specifically, we will give i) a value
$\alpha_M$ such that $\Lambda$ is non-increasing in
$(\alpha_M,\alpha_k)$ and ii) a function $u$ which is non-decreasing
in $[0,\alpha_M)$ and for which $\Lambda(\alpha,k,r) \le
u(\alpha,k,r)$. Thus, we will conclude $g(k,r)\le u(a_M,k,r)$.

We begin by getting an upper bound for $\Lambda'$, as follows:
\begin{eqnarray}
  \Lambda'(\a,k,r) & = & -\ln\a+\ln(1-\a)-r
	\frac{k(1-\a)^{k-1}}{2^k+(1-\a)^k-2} \nonumber \\ 
	& \le &  -\ln \a-\a-2^{-k} r k(1-\a)^{k-1} \nonumber \\
	& < &  -\ln \a-2^{-k} r k(1-\a)^{k-1} \label{opa}\\
	& \le & -\ln \a-2^{-k} r k(1-k\a) \nonumber \\
	& \equiv & \hat{u}(\a,k,r) \nonumber \enspace .
\end{eqnarray}
\begin{lemma}\label{fede1}
If $r=\gamma 2^k \ln 2$, then for all $k \ge 8$ and $\gamma > 3k^{-1}
\log_2 k$, there exists
\begin{equation}
\a_M \; \le \;   
2^{-\gamma k} (1 + 4 \g k^2 2^{-\g k} \ln 2)   \enspace , 
\label{eq:bounds_aM}
\end{equation}
such that $\hat{u}(\a_M,k,r)=0$. 
\end{lemma}
\begin{proof}[Proof of Lemma~\ref{fede1}.] Let
$$
q(\alpha) = 2^{-\gamma k} 2^{\gamma k^2 \alpha} \enspace .
$$
We begin by noting that if $\alpha_M$ is such that
$q(\alpha_M)=\alpha_M$ then $\hat{u}(\alpha_M,k,r)=0$.
Now, let us define
$$
s(\alpha) = 2^{-\g k} (1 + 2 \alpha \g k^2 \ln 2) \enspace .
$$
Observe that the unique solution of $s(\alpha) = \alpha$ is
\begin{equation}\label{astar}
\a^* = \frac{2^{-\g k}}{1 - 2 \g k^2 2^{-\g k} \ln 2} 
\end{equation}
and that $s(\alpha) > \alpha$ for all $\alpha \in [0,\alpha^*)$.

Recall that $e^x \le 1+2x$ for all $0 \le x\leq 1$. Therefore,
$q(\alpha) < s(\alpha)$ for all $\alpha$ such that $\gamma k^2 \alpha
\ln 2 \leq 1$. In particular, if $\g k^2 \alpha^* \ln 2 \le 1$, then
since $s(\alpha) > \alpha$ for all $\alpha \in [0,\alpha^*)$, we can
conclude that the equation $q(\alpha)=\alpha$ has at least one root
$\alpha_M \le \alpha^*$, as desired.

By~\eqref{astar}, the condition $\g k^2 \alpha^* \ln 2 \le 1$ is
equivalent to
\begin{equation}
\label{eq:condition}
\g k^2 2^{-\g k} \le
\frac{1}{3 \ln(2)} = 0.4808...
\end{equation}
To establish that~\eqref{eq:condition} holds we note that for any $\g
> 3 k^{-1} \log_2 k$ the quantity $\g k^2 2^{-\g k}$ is decreasing in
$\g$ and, therefore, it is bounded by $z(k)=3k^{-2}\log k$. As $z(k)$
is decreasing for $k\ge 2$, for all $k \ge 8$ we have $\g k^2 2^{-\g
k} \le z(8) = 9/64 = 0.1406... < 0.4808...$, as desired. The fact $\g
k^2 2^{-\g k} \le 0.1406...$ along with the inequality $1/(1-x) \le
1+2x$ valid for $x \le 1/2$, gives us $\alpha_M \le \alpha^* \le
{2^{-\g k}}(1 +4 \g k^2 2^{-\g k} \ln 2)$.
\end{proof}

To bound $\Lambda$ by an non-decreasing function we note
\begin{equation}
\ln \Lambda(\a,k,r) \le  \ln 2 - \a\ln \a  +\a-r 2^{-k} (1+\a) \equiv
u(\a,k,r)\enspace . \label{udef} 
\end{equation}
\begin{lemma}
If $r=\gamma 2^k \ln 2$, then for every $k \ge 8$ and $\gamma \in
(3k^{-1} \log_2 k,1]$,
$$
 u(a_M,k,r) \le (1-\g) \ln 2 + \left(1+\frac{9 \ln 2}{16} \right)
 k^{-2}  \enspace  . 
$$
\end{lemma} 
\begin{proof}
Using Lemma~\ref{fede1} to pass from~\eqref{kola} to~\eqref{mola}, we
see that for every $k \ge 8$ and $\gamma \in (3k^{-1} \log_2 k,1]$,
\begin{eqnarray}
 u(\a_M,k,r) & = & \ln 2 + \a_M \big(\gamma k \ln 2 - \gamma k^2
   \a_M \ln 2 \big) + \a_M - \gamma \ln 2 (1 + \a_M) \nonumber  \\ 
& \le &  (1-\gamma)\ln 2  + \a_M \big[1 + \gamma (k-1) \ln 2\big]
   \label{kola} \\ 
& \le &  (1-\gamma)\ln 2  + 2^{-\gamma k} (1 + 4  \g k^2 2^{-\g k} \ln
   2) (\gamma k \ln 2  + 1) \label{mola} \enspace .
\end{eqnarray}
Recalling that~\eqref{eq:condition} holds for all $k \ge 8$ and
$\gamma > 3k^{-1} \log_2 k$, we conclude
\begin{eqnarray}
 u(\a_M,k,r) 
& \le &  (1-\gamma)\ln 2 + k^{-3} \left(1+\frac{9 \ln 2}{16} \right)
 (k \ln 2 + 1) \nonumber \\ 
& \le &  (1-\gamma)\ln 2   + \left(1+\frac{9\ln 2}{16} \right) k^{-2}
 \nonumber\enspace . 
\end{eqnarray} 
\end{proof} 

We can now prove Lemma~\ref{lemvol}.
\begin{proof}[Proof of Lemma~\ref{lemvol}.]
Recall the definition of the function $u$ from~\eqref{udef} and note
that, since $u'(\alpha)=-\ln \alpha -r2^{-k}$, it is non-decreasing
for $r \le 2^k$ and $\alpha \le 1/e$. From~\eqref{eq:bounds_aM} we see
that $\alpha_M < 1/e$ and therefore we can conclude that
$\Lambda(\alpha,k,r) < u(\a_M,k,r)$ for all $\alpha \in
[0,\alpha_M)$. To complete the proof it thus suffices to prove that
$\Lambda$ is non-increasing in the interval $(\alpha_M,1/k)$ since, by
our results in the previous section, we know that $\Makis \le 1/k$
both when $\gamma \ge 49/50$ and $k>11$, and when $\gamma \in (2/3,1)$
and $k>15$.  For that we first observe that
$$
\hat{u}'(\alpha,k,r) = - \frac{1}{\alpha}+2^{-k}rk^2 < -
\frac{1}{\alpha} + k^2 \enspace .
$$
Since, by definition, $\hat{u}(\alpha_M,k,r)=0$ this implies
$\hat{u}\le 0$ for all $\alpha \in [\a_M, 1/k^2]$ and since $\Lambda'
\le \hat{u}$, it follows that $\Lambda' \le 0$ also for such
$\alpha$. Using~\eqref{opa}, it is straightforward to check that for
$\alpha \in [1/k^2, 1/k]$, the derivative of $\Lambda$ is negative
both when i) $\gamma \ge 49/50$ and $k>11$, and when ii) $2/3 <\gamma
< 1$ and $k>15$, thus concluding the proof.
\end{proof}
 
\subsection{Proof of Lemma~\ref{lembal}: A lower bound on the number
  of balanced assignments}

\begin{proof}[Proof of Lemma~\ref{lembal}.]
Recalling the definition of $\Lambda_b$ from~\cite{AP} we have 
\begin{equation}
\label{eq:b}
\ln \Lambda_b(1/2,k,r) = 2 \ln 2 + r
\ln\left[\frac{\big((1-\epsilon/2)^k-2^{-k}\big)^2}{(1-\epsilon)^k}\right]
\enspace ,
\end{equation}
where $\epsilon$ satisfies
\begin{equation}\label{epdef}
\epsilon (2-\epsilon)^{k-1} =1 \enspace . 
\end{equation} 
We note for later use that, as shown in~\cite{AP}, if $\epsilon$
satisfies~\eqref{epdef} then
\begin{equation}\label{epbounds}
2^{1-k} + k4^{-k} < \epsilon < 2^{1-k} + 3k4^{-k} \enspace .
\end{equation}

Since all coefficients in the binomial expansion of
$(1-\epsilon)^{-k}$ are positive,
\begin{equation}\label{omana}
(1-\epsilon)^{-k} \ge 1 + k \epsilon + \frac{k(k+1)}{2} \epsilon^2 \enspace .
\end{equation}
To get a lower bound for the numerator inside the logarithm
in~\eqref{eq:b} we consider the binomial expansion of
$(1-\epsilon/2)^k $. We observe that the sum of a pair of successive
terms where the lower term corresponds to an even power equals
\begin{equation}\label{last}
\binom{k}{j}(\epsilon/2)^j - \binom{k}{j+1}(\epsilon/2)^{j+1} = 
\binom{k}{j}(\epsilon/2)^j \left[1 -
  \frac{(k-j)\epsilon}{2(j+1)}\right] \enspace .
\end{equation}
For $k \ge 8, j \ge 4$ and $\epsilon \le 5/2$ the expression
in~\eqref{last} is positive. Moreover, when $k$ is even the last term
in the binomial expansion has a positive coefficient and can be safely
discarded. Therefore, for all $k \ge 8$ and $\epsilon \le 5/2$,
\begin{equation}\label{opapa}
(1-\epsilon/2)^k 
\ge 1 - \frac{k\epsilon}{2} + \frac{k(k-1)\epsilon^2}{8} -
\frac{k(k-1)(k-2)\epsilon^3}{48} \enspace .
\end{equation}

Substituting~\eqref{omana} and~\eqref{opapa} into~\eqref{eq:b} we get
a lower bound of the form $\ln \Lambda_b \ge c_0 +c_1 \epsilon+ c_2
\epsilon^2 \cdots + c_8 \epsilon^8$. It is not hard to check directly
that $c_8 \ge 0$ for all $k\ge 8$. Similarly, using the upper bound
for $\epsilon$ from~\eqref{epbounds}, it is not hard to check that for
$i=2,4,6$, we have $c_i + c_{i+1} \epsilon \geq 0$ for all $k \ge
8$. Therefore, we can conclude
\begin{eqnarray}
  \ln \Lambda_b(1/2,k,r) & \ge & 2 \ln 2 + r
  \ln\left[1 - 2^{1-k} + 2^{-2k} - \epsilon k 2^{-k} (1-2^{-k})
  \right] \nonumber \\
  & \ge & 2 \ln 2  + r \ln\left[1 - 2^{1-k} + 2^{-2k} -
    k 2^{-k} (1-2^{-k}) (2^{1-k} + 3 k 2^{-2k}) \right] \enspace ,
  \label{fola}
\end{eqnarray}
where in~\eqref{fola} we have replaced $\epsilon$ with its upper bound
from~\eqref{epbounds}.

The argument of the logarithm in~\eqref{fola} is increasing in $k$ for
all $k \ge 3$ (a fact that can be easily established by considering
its derivative). As a result, we have that for all $k \geq 8$, it is
at least equal to its value for $k=8$ which is $1-0.00805183... >
1/2$. Thus, using the inequality $\ln(1+x) > x - x^2$ valid for all $x
> -1/2$, we can finally write
$$
\ln \Lambda_b(1/2,k,\gamma 2^k \ln 2) \ge 2 \ln 2 [1 -  \gamma m(k)]
\enspace ,
$$ 
where
\begin{multline}
m(k) = 1 + \frac{2k+3}{2} 2^{-k} + \frac{3k^2+6k-4}{2} 2^{-2k} +
\frac{13k^2-12k+1}{2} 2^{-3k} \\
+ (6k^3-13k^2+2k) 2^{-4k}+ \frac{9k^4-24k^3+10k^2}{2} 2^{-5k} +
(9k^4-6k^3) 2^{-6k} + \frac{9}{2} k^4 2^{-7k}
\end{multline}
\end{proof}

\section*{Acknowledgements}

This work has been partially supported by the EC through the FP6 IST
integrated project ``EVERGROW''.

\end{document}